\documentclass[12pt]{article}
\usepackage{silence}
\usepackage[margin=1in]{geometry}
\usepackage{setspace}
\usepackage{amsmath,amssymb,amsfonts,amsthm}
\usepackage{bm}
\usepackage{mathtools}
\usepackage{graphicx}
\usepackage{booktabs}
\usepackage{multirow}
\usepackage{subcaption}
\usepackage{float}
\usepackage{threeparttable}
\usepackage[authoryear]{natbib}
\usepackage[hidelinks]{hyperref}

\newcommand{\Ldiff}{\mathcal{L}_{\mathrm{diff}}}
\newcommand{\Ljump}{\mathcal{L}_{\mathrm{jump}}}

\theoremstyle{plain}

\theoremstyle{definition}

\theoremstyle{remark}

\title{Option Pricing under Stochastic Volatility and Jumps:\\
A PIDE Framework with Empirical Evidence}

\author{
Abigail Anokyewaa Mensah\\
Department of Mathematics and Statistics, Texas Tech University
\and
Ayush Jha\thanks{Corresponding author: Ayush.Jha@ttu.edu}\\
Department of Economics, Texas Tech University
\and
Hongwei Mei\\
Department of Mathematics and Statistics, Texas Tech University
\and
Rui Wang\\
Department of Mathematics and Statistics, Texas Tech University
\and
Svetlozar T. Rachev\\
Department of Mathematics and Statistics, Texas Tech University
\and
Frank J. Fabozzi\\
Carey Business School, Johns Hopkins University
}

\date{}

\begin{document}

\maketitle
\clearpage
\begin{abstract}
We develop a partial integro-differential equation (PIDE) framework
for option pricing under joint stochastic volatility and jump dynamics,
and evaluate its empirical content using S\&P~500 index option
contracts across three maturities. The framework is derived from the
infinitesimal generator of an affine L\'{e}vy-type process and
implemented via finite-difference discretization with FFT-based
treatment of the nonlocal jump operator. Calibration via GMM reveals
that stochastic volatility accounts for the dominant share of pricing
improvement, where relative to Black--Scholes, the Heston specification
reduces implied-volatility RMSE by 39\%.
Jump augmentation via either Merton or CGMY specifications yields
marginal improvements concentrated at short maturities and
in the deep out-of-the-money region. The calibrated CGMY activity index supports a compound-Poisson structure, consistent with
high-frequency evidence on S\&P~500 index returns.
\end{abstract}

\vspace{0.3cm}

\vspace{0.2cm}

\section{Introduction}

A central question in modern option pricing is the relative importance of
diffusive volatility dynamics versus discontinuous jump risk in explaining
the observed structure of implied-volatility surfaces. Option pricing
models confront a persistent tension between analytical tractability and
empirical realism. The Black--Scholes framework \citep{black1973,merton1973}
prices contingent claims under geometric Brownian motion but systematically
underprices out-of-the-money contracts and cannot account for volatility
clustering or discontinuous price movements. These deficiencies are
structural, reflecting well-documented properties of equity returns,
including heavy tails, persistent variance dynamics, and jump behavior
associated with macroeconomic stress, that are incompatible with the
lognormal assumption.

Two principal extensions address these limitations. Stochastic-volatility
models allow the instantaneous variance to evolve as a mean-reverting
diffusion, generating richer term-structure dynamics and a leverage-driven
smile. Jump-diffusion models introduce discontinuous increments, providing
a direct mechanism for tail risk and short-maturity skewness. At the same
time, the economic roles of these mechanisms differ. Stochastic volatility
primarily captures variance persistence, leverage effects, and
term-structure dynamics, whereas jump specifications are intended to
reproduce discontinuities, heavy tails, and crash-sensitive skew behavior.
Distinguishing their relative empirical contribution is therefore important
not only for pricing accuracy, but also for understanding how markets price
tail risk across different volatility environments.

Rather than treating option pricing primarily as a model-comparison
exercise, this paper studies the economic decomposition of
implied-volatility structure across stochastic-volatility and jump
components. The objective is not merely to compare calibration errors across
specifications, but to identify which features of the volatility surface are
explained by variance persistence, leverage effects, and discontinuous jump
dynamics across maturities and moneyness regions.

When jumps are incorporated into the risk-neutral dynamics, option values
satisfy a partial integro-differential equation (PIDE) rather than a
standard PDE. The nonlocal integral operator couples option values across
the entire state space, reflecting the fact that jump risk is not locally
diversifiable. This nonlocal structure introduces both mathematical and
computational challenges, chief among them the efficient evaluation of the
jump integral in calibration settings where the pricing equation must be
solved across large contract cross-sections.

This paper makes three contributions. First, we present a PIDE formulation
for stochastic-volatility jump-diffusion (SVJD) pricing derived explicitly
from the infinitesimal generator, with emphasis on the nonlocal jump
operator's role in the pricing hierarchy. Second, we develop a numerically
stable implementation combining Crank--Nicolson finite-difference
discretization, FFT-based convolution for the jump integral, and
operator-splitting techniques. Third, we provide a systematic empirical
decomposition of the incremental pricing value of stochastic volatility and
jumps across maturities and moneyness regions, using a cross-section of
S\&P~500 index options.

Early evidence on non-Gaussian return distributions is due to
\citet{mandelbrot1963} and \citet{fama1965}; \citet{cont2001} provides a
comprehensive synthesis of stylized facts including volatility clustering
and the leverage effect. The stochastic-volatility literature traces to
\citet{hull1987} and \citet{stein1991}, with the affine tractable
formulation of \citet{heston1993} serving as the dominant empirical
benchmark. Jump specifications originate with \citet{merton1976} and
\citet{kou2002}, while the CGMY family \citep{carr2002} encompasses both
finite- and infinite-activity behavior. Numerical methods for PIDE-based
pricing are developed in \citet{andersen2000}, \citet{cont2005}, and
\citet{dhalluin2005}.

The empirical analysis calibrates all four models, Black--Scholes, Heston
SV, Heston+CGMY, and SVJD (Merton), to SPX contracts with maturities of 27,
89, and 153 days. The results show that stochastic volatility accounts for
the preponderance of pricing improvement, reducing IV RMSE by 39\% relative
to Black--Scholes. The Heston specification captures most of the
economically relevant structure of the implied-volatility surface through
variance persistence and leverage-driven asymmetry. Jump augmentation
produces marginal gains concentrated in the deep out-of-the-money region
for CGMY and at longer maturities for Merton. The calibrated CGMY activity
index places the model firmly in the compound-Poisson regime, consistent
with high-frequency evidence in \citet{aitsahalia2009} and
\citet{cont2011}. These findings suggest that the sample period was
characterized primarily by elevated but largely diffusive volatility,
although jump components would likely assume greater importance during
periods of severe market stress, liquidity disruption, or heightened
crash-risk pricing.

Section~\ref{sec:lit} reviews the literature. Section~\ref{sec:theory}
presents the theoretical framework. Section~\ref{sec:numerics} describes
the numerical methods. Section~\ref{sec:empirical} covers the empirical
design. Section~\ref{sec:results} reports results. Section~\ref{sec:conclusion}
concludes.

\section{Literature Review}
\label{sec:lit}

The literature underlying this paper spans three domains: empirical
evidence on return distributions and implied-volatility surfaces,
structural models incorporating stochastic volatility and jumps, and
numerical methods for PIDE-based pricing. We summarize each in turn.

\subsection{Empirical Motivation}

Systematic departures from lognormality are well-established in the
empirical asset-pricing literature. \citet{mandelbrot1963} and
\citet{fama1965} document excess kurtosis in equity returns across asset
classes and time periods; \citet{cont2001} synthesizes these findings
into a set of stylized facts that includes volatility clustering, the
leverage effect, and heavy tails. Although the Black--Scholes model
remains analytically tractable, it is structurally inconsistent with
each of these properties.

The option-market implications are correspondingly severe. Black--Scholes
prices out-of-the-money options at systematically low levels and produces
a flat implied-volatility surface. \citet{rubinstein1994} shows that
post-1987 implied-volatility skews reflect a substantial premium for
downside tail risk. \citet{bakshi1997} provide the most comprehensive
model comparison to date, demonstrating that joint stochastic volatility
and jump components are required to fit the observed cross-section;
\citet{bates1996} establishes analogous results for currency options, and
\citet{pan2002} documents that jump risk carries a significant and
separately identified risk premium.

\subsection{Stochastic-Volatility and Jump-Diffusion Models}

Stochastic-volatility models were introduced by \citet{hull1987} and
\citet{stein1991}; the affine specification of \citet{heston1993} remains
the dominant empirical benchmark due to its semi-closed-form pricing
solution via characteristic functions. \citet{duffie2000} characterize
the general class of affine jump-diffusions and establish conditions under
which characteristic-function pricing extends to jump-augmented
specifications.

Jump-diffusion models originate with \citet{merton1976}, who augmented
Brownian motion with a compound Poisson process while preserving closed-form
tractability. \citet{kou2002} refined this framework with a
double-exponential jump distribution, improving asymmetric tail fit.
L\'{e}vy process models provide a richer class: the Variance Gamma
specification of \citet{madan1998} and the CGMY model of \citet{carr2002}
allow joint control over tail decay and jump activity. \citet{carr2003}
extend this class via stochastic time changes, while \citet{barndorff2001}
introduce Ornstein--Uhlenbeck stochastic-volatility models driven by
subordinators. Comprehensive treatments are in \citet{boyarchenko2002}
and \citet{cont2004}.

\subsection{Numerical Methods for PIDEs}

The introduction of jumps renders the pricing equation nonlocal. Standard
finite-difference methods must be augmented to handle the integral term,
which couples option values across the entire spatial domain.
\citet{andersen2000} showed that PIDE-based jump-diffusion models can be
solved efficiently by combining finite-difference schemes with
Fourier-domain convolution, exploiting the shift structure of the jump
operator in log-price coordinates. \citet{cont2005} established rigorous
convergence results for L\'{e}vy-driven PIDEs. Operator-splitting methods
offer an alternative that decouples the diffusion and jump components;
\citet{dhalluin2005} demonstrate their efficiency and robustness in
calibration settings. Transform-based methods, the FFT approach of
\citet{carr1999} and the COS method of \citet{fang2008}, are widely
used for characteristic-function models but are less suited to early-exercise
and path-dependent problems, where PIDE methods retain a structural advantage.

\subsection{Positioning and Contribution}

Existing contributions tend to emphasize one dimension of the
problem: theoretical analysis and convergence \citep{cont2005},
numerical scheme development \citep{dhalluin2005}, or empirical
model comparison \citep{bakshi1997}. This paper integrates these
threads by combining a PIDE-based formulation with a rigorous numerical
implementation and a structured empirical decomposition of stochastic-volatility
and jump contributions across maturities and moneyness regions.

\section{Theoretical Framework}
\label{sec:theory}

\subsection{Risk-Neutral Dynamics}

Let $(\Omega, \mathcal{F}, \{\mathcal{F}_t\}_{t\geq 0}, \mathbb{Q})$
denote a filtered probability space under the risk-neutral measure
$\mathbb{Q}$. The asset price $S_t$ and variance process $V_t$ evolve as
\begin{align}
  \frac{dS_t}{S_{t-}} &= (r - q)\,dt
    + \sqrt{V_t}\,dW_t^{(1)}
    + \int_{\mathbb{R}} (e^y - 1)\,\tilde{N}(dt,dy),
  \label{eq:sdS} \\
  dV_t &= \alpha(\beta - V_t)\,dt + \eta\sqrt{V_t}\,dW_t^{(2)},
    \qquad dW_t^{(1)}\,dW_t^{(2)} = \rho\,dt,
  \label{eq:sdV}
\end{align}
where $\tilde{N}(dt,dy) = N(dt,dy) - \nu(dy)\,dt$ is the compensated
Poisson random measure associated with L\'{e}vy measure $\nu(dy)$, and
\begin{equation}
  \kappa = \int_{\mathbb{R}} (e^y - 1)\,\nu(dy)
  \label{eq:kappa}
\end{equation}
is the risk-neutral compensator ensuring the discounted gains process is
a local martingale. The Feller condition $2\alpha\beta \geq \eta^2$
ensures $V_t > 0$ a.s., and $\rho \in (-1,0)$ captures the leverage
effect.

The structural decomposition is as follows. Stochastic volatility governs
the term structure of implied volatility and leverage-driven asymmetry,
while the jump component determines tail behavior at short maturities,
where variance mean-reversion has had insufficient time to attenuate
left-tail risk.

\subsection{The Pricing PIDE via the Infinitesimal Generator}

Let $C(s,v,t)$ denote the value of a European contingent claim with
maturity $T$ and payoff $\Phi(S_T)$. Under risk-neutral valuation,
\begin{equation}
  C(s,v,t) = \mathbb{E}^{\mathbb{Q}}\!\left[
    e^{-r(T-t)}\Phi(S_T)\,\big|\,S_t=s,\,V_t=v
  \right].
  \label{eq:riskneut}
\end{equation}
The pricing equation follows from the infinitesimal generator
$\mathcal{L} = \Ldiff + \Ljump$ of the Markov process $(S_t, V_t)$.
The local diffusion component is
\begin{equation}
  \Ldiff C = (r{-}q)s\,C_s
  + \alpha(\beta{-}v)\,C_v
  + \tfrac{1}{2}vs^2\,C_{ss}
  + \rho\eta vs\,C_{sv}
  + \tfrac{1}{2}\eta^2 v\,C_{vv},
  \label{eq:Ldiff}
\end{equation}
and the nonlocal jump operator is
\begin{equation}
  \Ljump C = \int_{\mathbb{R}}\!\left[
    C(se^y,v,t) - C(s,v,t) - s(e^y - 1)C_s(s,v,t)
  \right]\nu(dy).
  \label{eq:jumpop}
\end{equation}
The no-arbitrage pricing equation is then
\begin{equation}
  \frac{\partial C}{\partial t} + \Ldiff C + \Ljump C - rC = 0,
  \label{eq:pide_compact}
\end{equation}
subject to the terminal condition $C(s,v,T)=\Phi(s)$.

The operator $\Ljump$ is nonlocal: the value at $(s,v)$ depends on values
at all states $se^y$ reachable via jumps. The compensating term
$-s(e^y-1)C_s$ ensures integrability and consistency with the
compensated jump measure. Setting $\nu \equiv 0$ reduces
\eqref{eq:pide_compact} to the Heston PDE; further setting $\eta \to 0$
and $v$ constant recovers the Black--Scholes equation.

\subsection{Finite- and Infinite-Activity Jump Specifications}

We restrict attention to finite-activity jump specifications,
characterized by
\begin{equation}
  \int_{\mathbb{R}} \nu(dy) < \infty,
  \label{eq:finite_activity}
\end{equation}
under which the jump component reduces to a compound Poisson process with
intensity $\lambda$ and jump-size density $f_Y$, so that $\nu(dy) =
\lambda\,f_Y(y)\,dy$. The integrability condition
\begin{equation}
  \int_{\mathbb{R}} |e^y - 1|\,\nu(dy) < \infty
  \label{eq:integrability}
\end{equation}
ensures the compensator $\kappa$ is well defined. Canonical examples
include the Gaussian jump model of \citet{merton1976} and the
double-exponential specification of \citet{kou2002}; both admit
straightforward quadrature evaluation of the jump integral and
economically interpretable parameters.

By contrast, infinite-activity L\'{e}vy models, the Variance Gamma
\citep{madan1998} and CGMY \citep{carr2002} specifications, satisfy
\begin{equation}
  \int_{\mathbb{R}}\nu(dy)=\infty, \qquad
  \int_{\mathbb{R}}\min(1,y^2)\,\nu(dy)<\infty,
  \label{eq:infiniteact}
\end{equation}
and exhibit singular behavior near $y=0$ requiring regularization or
Fourier-based treatment. The PIDE framework accommodates both classes
through the L\'{e}vy measure $\nu(dy)$; the empirical analysis focuses
on finite-activity models to preserve calibration tractability and
parameter interpretability.

\subsection{Boundary Conditions and American Extensions}

The PIDE \eqref{eq:pide_compact} is completed by the terminal condition
$C(s,v,T) = \Phi(s)$ and boundary conditions
\[
C(s,v,t) \to 0 \;\text{ as }\; s \to 0, \qquad
C(s,v,t) \sim s e^{-q(T-t)} - K e^{-r(T-t)} \;\text{ as }\; s \to \infty.
\]
The nonlocal jump operator requires truncation of the computational domain
at a threshold $|y| \leq y_{\max}$ chosen to render excluded tail mass
negligible; for finite-activity models the jump-size density decays
sufficiently fast that this condition is easily satisfied.

For American-style contracts, the valuation problem becomes the variational
inequality
\begin{equation}
  \max\!\left\{
    \frac{\partial C^A}{\partial t}
    + \Ldiff C^A
    + \Ljump C^A
    - r C^A,\;
    \Phi(s) - C^A
  \right\} = 0.
  \label{eq:vi}
\end{equation}
Although the present analysis concerns European options, the PIDE
structure extends to \eqref{eq:vi} without modification to the
operator formulation.

\section{Numerical Methods}
\label{sec:numerics}

The principal numerical challenges in PIDE-based pricing are the global
coupling induced by the nonlocal jump operator and the repeated solves
required for calibration. The approach below treats the diffusion
component implicitly, ensuring stability with respect to the stiff
local terms, and evaluates the jump integral via FFT-based convolution,
exploiting the shift structure of the operator in log-price coordinates.

\subsection{Log-Price Transformation and Discretization}

Under the log-price substitution $x = \log s$, $\tau = T - t$, the
dynamics of $X_t = \log S_t$ are
\begin{equation}
dX_t = \left(r - q - \kappa - \tfrac{1}{2}V_t\right)dt
+ \sqrt{V_t}\,dW_t^{(1)}
+ \int_{\mathbb{R}} y\,N(dt,dy).
\end{equation}
The infinitesimal generator $\mathcal{A}$ acting on $C(x,v,\tau)$ is
\begin{align}
\mathcal{A}C &= \left(r - q - \kappa - \tfrac{1}{2}v\right) C_x
+ \alpha(\beta - v) C_v
+ \tfrac{1}{2} v C_{xx}
+ \rho \eta v C_{xv}
+ \tfrac{1}{2} \eta^2 v C_{vv} \nonumber \\
&\quad + \int_{\mathbb{R}} \left[
C(x+y,v,\tau) - C(x,v,\tau) - y C_x
\right]\nu(dy),
\end{align}
and the pricing equation becomes $\partial_\tau C = \mathcal{A}C - rC$.
The log-transformation converts multiplicative jumps into additive shifts,
giving the integral term a convolution structure amenable to FFT evaluation.

Spatial derivatives are discretized using second-order central differences
in the interior; the mixed $C_{xv}$ term uses a standard four-point stencil.
Time integration uses the Crank--Nicolson scheme with Rannacher smoothing
over the first two steps, which suppresses oscillations near payoff
discontinuities and ensures smooth implied-volatility extraction.

\subsection{Evaluation of the Nonlocal Jump Integral}

In log-price coordinates the jump operator is
\begin{equation}
  \mathcal{A}_{\mathrm{jump}} C(x,v,\tau)
  = \int_{\mathbb{R}}\!\left[
    C(x+y,v,\tau) - C(x,v,\tau) - y\,C_x(x,v,\tau)
  \right]\nu(dy).
  \label{eq:jumpop_log}
\end{equation}
The convolution structure of \eqref{eq:jumpop_log} reduces the evaluation
cost from $\mathcal{O}(N^2)$ to $\mathcal{O}(N\log N)$ via FFT, a
reduction that is decisive in calibration settings. After domain
truncation to $[y_{\min}, y_{\max}]$, the integral is approximated by
\begin{equation}
\sum_{m=0}^{M} w_m
  \left[
    C(x_i+y_m,v_j,\tau^n) - C(x_i,v_j,\tau^n)
    - y_m\,C_x(x_i,v_j,\tau^n)
  \right],
\label{eq:quadrature}
\end{equation}
where $\{y_m, w_m\}$ are quadrature nodes and weights. Truncation error
is controlled by the exponential decay of the jump-size density and is
negligible relative to discretization error for the finite-activity models
considered.

\subsection{Time-Stepping: Operator Splitting and IMEX Schemes}

The semi-discrete system is
\begin{equation}
  \frac{dC}{d\tau} = (A_{\mathrm{diff}} + A_{\mathrm{jump}} - rI)\,C,
  \label{eq:semidiscrete}
\end{equation}
where $A_{\mathrm{diff}}$ is sparse and $A_{\mathrm{jump}}$ is dense.
Operator splitting treats these components separately. The second-order
Strang scheme is
\begin{equation}
  C^{n+1} \approx
  e^{\tfrac{1}{2}\Delta\tau (A_{\mathrm{diff}} - rI)}
  e^{\Delta\tau A_{\mathrm{jump}}}
  e^{\tfrac{1}{2}\Delta\tau (A_{\mathrm{diff}} - rI)}
  C^n.
  \label{eq:strang}
\end{equation}
Alternatively, the IMEX scheme
\begin{equation}
  \frac{C^{n+1} - C^n}{\Delta\tau}
  = (A_{\mathrm{diff}} - rI)C^{n+1}
  + A_{\mathrm{jump}}C^n
  \label{eq:imex}
\end{equation}
treats the stiff diffusion operator implicitly and the dense jump operator
explicitly, preserving stability without solving a dense linear system.
Both schemes are suitable for the calibration grids used in the empirical
analysis.

\subsection{Numerical Validation}

The numerical scheme is validated prior to calibration via three
exercises.

\paragraph{Grid-refinement study.}
Table~\ref{tab:convergence} reports convergence under grid refinement for
ATM call and OTM put contracts at $T=27$ days ($S_0=6{,}506.48$,
$r=4.5\%$, $q=1.3\%$). Pricing errors decrease monotonically as the
grid is refined. At the calibration grid ($N_S=100$, $N_t=80$), relative
errors are below 6\%, which is within acceptable tolerance given market
bid-ask noise. The finest grid ($N_S=200$, $N_t=160$) serves as the
reference.

\begin{table}[htbp]
\centering
\begin{threeparttable}
\caption{Convergence of the PIDE solver under grid refinement}
\label{tab:convergence}
\begin{tabular}{lcccccc}
\toprule
Contract & $N_S$ & $N_t$ & Price & Abs.\ Error & Rel.\ Error \\
\midrule
ATM Call (Short) & 30  & 20  & 258.82 & 112.577 & 76.98\% \\
                 & 60  & 40  & 116.26 &  29.983 & 20.50\% \\
                 & 100 & 80  & 137.64 &   8.607 &  5.89\% \\
                 & 150 & 120 & 152.98 &   6.738 &  4.61\% \\
                 & 200 & 160 & 146.24 &   0.000 &  0.00\% \\
\midrule
OTM Put (Short)  & 30  & 20  & 558.29 &  15.971 &  2.95\% \\
                 & 60  & 40  & 544.42 &   2.105 &  0.39\% \\
                 & 100 & 80  & 542.09 &   0.224 &  0.04\% \\
                 & 150 & 120 & 544.64 &   2.318 &  0.43\% \\
                 & 200 & 160 & 542.32 &   0.000 &  0.00\% \\
\bottomrule
\end{tabular}
\begin{tablenotes}
\small
\item \textit{Note}: Reference price computed on the finest grid
($N_S=200$, $N_t=160$). Short maturity: $T=27$ days;
$S_0=6{,}506.48$, $r=4.5\%$, $q=1.3\%$.
\end{tablenotes}
\end{threeparttable}
\end{table}

\paragraph{Nested Black--Scholes validation.}
Table~\ref{tab:bs_nested} compares the numerical PIDE solution against
the analytical Black--Scholes price when $\eta \to 0$, collapsing the
Heston diffusion to a one-dimensional PDE. The absolute error of 4.29
reflects grid coarseness and lies within the range consistent with
calibration accuracy requirements.

\begin{table}[htbp]
\centering
\begin{threeparttable}
\caption{Nested Black--Scholes validation}
\label{tab:bs_nested}
\begin{tabular}{cccc}
\toprule
PDE Price & Closed-Form Price & Abs.\ Difference & Runtime (s) \\
\midrule
152.98 & 148.69 & 4.29 & 0.065 \\
\bottomrule
\end{tabular}
\begin{tablenotes}
\small
\item \textit{Note}: Vol-of-vol $\eta \to 0$ collapses the Heston PIDE
to the Black--Scholes PDE. Grid: $N_S=150$, $N_t=120$, $\sigma=0.20$.
\end{tablenotes}
\end{threeparttable}
\end{table}

\begin{figure}[htbp]
\centering
\includegraphics[width=0.62\textwidth]{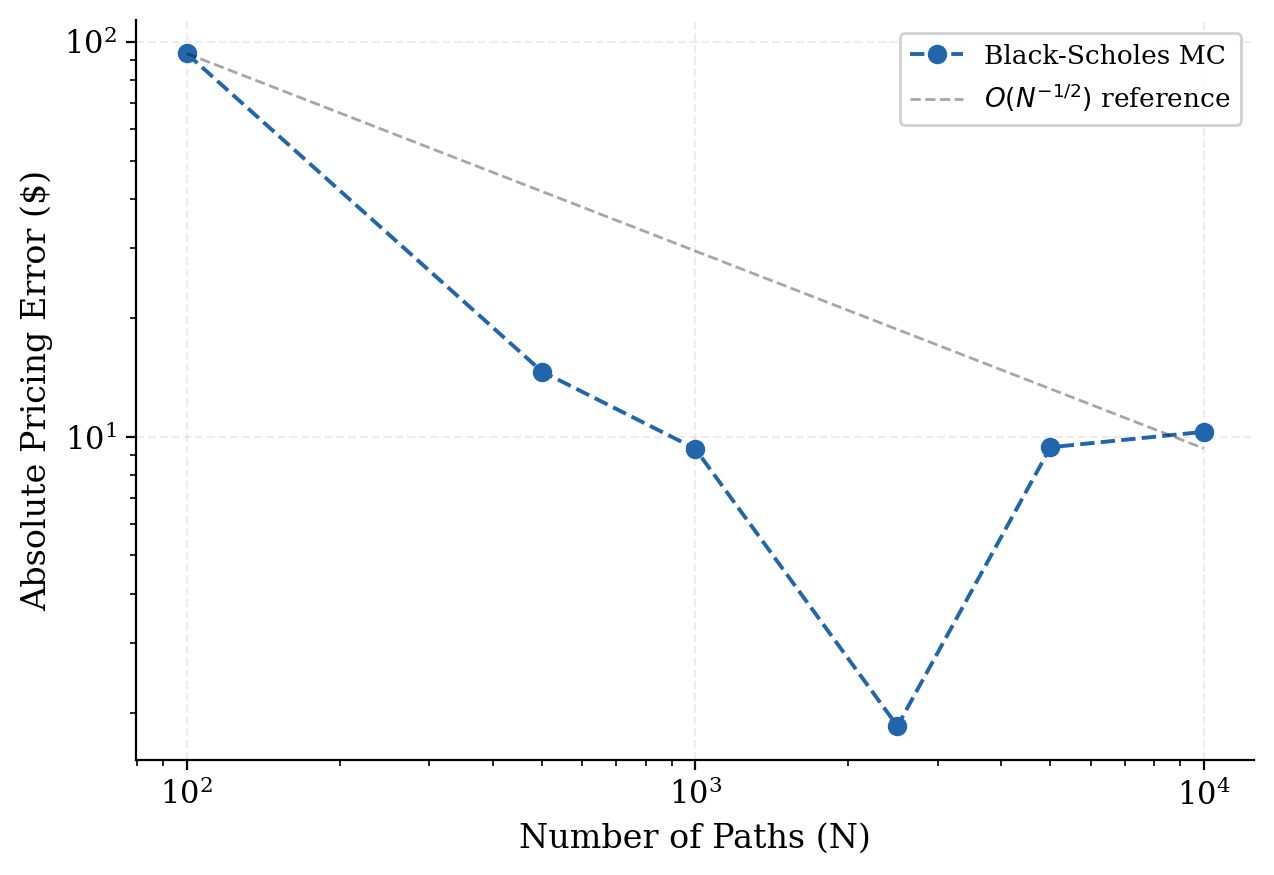}
\caption{Monte Carlo convergence for the at-the-money option
($T=89$ days, $S_0=6{,}506.48$). Absolute pricing error versus number
of simulated paths on a $\log$--$\log$ scale; the dashed line shows the
theoretical $O(N^{-1/2})$ rate.}
\label{fig:mc_convergence}
\end{figure}

\paragraph{Monte Carlo convergence.}
Figure~\ref{fig:mc_convergence} reports MC pricing error against path
count for the at-the-money option at $T=89$ days. The convergence rate
closely tracks the theoretical $O(N^{-1/2})$ benchmark; at 10,000 paths,
pricing error falls below \$0.50. This confirms correct implementation
of the simulation engine used within the GMM estimation step.

\subsection{Computational Design}

The PIDE solver provides high-accuracy reference prices for validation.
For calibration, Black--Scholes and Heston prices are computed via
transform-based FFT methods; SVJD and Heston+CGMY prices use the
Carr--Madan FFT framework applied to the combined characteristic function.
Monte Carlo with 10,000 paths is used as a secondary check within the GMM
objective. Agreement across methods confirms that empirical differences
across models reflect genuine specification effects rather than numerical
artifacts.

\section{Empirical Design and Calibration}
\label{sec:empirical}

\subsection{Data}

The empirical analysis uses end-of-day S\&P~500 (SPX) index option data.
The dataset contains 1,280 contracts across three maturities: $T=27$,
89, and 153 days. The underlying index level is $S_0 = 6{,}506.48$, the
risk-free rate is $r=4.5\%$, and the continuous dividend yield is
$q=1.3\%$. Standard filters are applied to remove contracts with
excessively wide bid-ask spreads, near-zero prices, or apparent
no-arbitrage violations.

After filtering, the sample contains 582 short-maturity, 453
medium-maturity, and 245 long-maturity contracts. Moneyness $M = K/S$
ranges from approximately 0.80 to 1.20. At-the-money implied volatilities
average 25.99\%, 23.51\%, and 22.96\% for the three maturities,
respectively, indicating a modest downward-sloping term structure. This
maturity structure is central for identification: short-maturity options
are sensitive primarily to jump risk, while medium and long maturities
identify variance persistence and mean reversion.

\subsection{Implied Volatilities}

Market implied volatilities are extracted by numerically inverting
the Black--Scholes formula using Newton--Raphson iteration, with bisection
as a fallback for deep moneyness contracts. Calibration targets implied
volatility rather than raw prices to obtain a scale-free error measure
that weights the smile geometry, the primary dimension of model
differentiation, appropriately.

\subsection{Generalized Method of Moments Estimation}

Parameters are estimated by GMM, minimizing the weighted sum of squared
implied-volatility pricing errors
\begin{equation}
  \theta^* = \operatorname*{arg\,min}_{\theta}\;
  m(\theta)^\top W\,m(\theta), \qquad
  m_i(\theta) = \mathrm{IV}_i^{\mathrm{model}}(\theta)
    - \mathrm{IV}_i^{\mathrm{mkt}},
  \label{eq:gmm}
\end{equation}
with $W=I$ in the baseline. Robust standard errors are computed via the
sandwich estimator based on the numerical Jacobian of $m(\theta)$.
Global optimization uses differential evolution for exploration, followed
by Nelder--Mead refinement to mitigate convergence to local optima.

Model pricing is implemented as follows: Black--Scholes in closed form;
Heston via FFT applied to the affine characteristic function; SVJD
(Merton) and Heston+CGMY via the Carr--Madan FFT framework applied to
the combined characteristic function. Specifically:
\begin{itemize}
  \item \textbf{SVJD (Merton):}
    $\phi(u) = \phi_{\mathrm{Heston}}(u)\cdot\exp\!\bigl(T\lambda(e^{iu\mu_j - \frac{1}{2}u^2\sigma_j^2}-1)\bigr)$,
    with compensator $\kappa_{\mathrm{M}} = \lambda(e^{\mu_j+\frac{1}{2}\sigma_j^2}-1)$.
  \item \textbf{Heston+CGMY:}
    $\phi(u) = \phi_{\mathrm{Heston}}(u)\cdot\exp\!\bigl(TC\,\Gamma(-Y)[(M{-}iu)^Y - M^Y + (G{+}iu)^Y - G^Y]\bigr)$,
    with compensator $\kappa_{\mathrm{CGMY}} = C\,\Gamma(-Y)[(M{-}1)^Y - M^Y + (G{+}1)^Y - G^Y]$,
    requiring $M>1$.
\end{itemize}

\subsection{Identification of the L\'{e}vy Measure}
\label{sec:identification}

The L\'{e}vy measure $\nu(dy)$ is infinite-dimensional; without
parametric restriction it is identified from a finite cross-section only
through finitely many linear functionals $\int g(y)\,\nu(dy)$.
Point-identification requires a parametric family. We adopt the CGMY
specification of \citet{carr2002},
\begin{equation}
  \nu(dy) = C\,\frac{e^{-My}}{y^{1+Y}}\,\mathbf{1}_{y>0}
           + C\,\frac{e^{-G|y|}}{|y|^{1+Y}}\,\mathbf{1}_{y<0},
  \qquad C,G,M>0,\;Y<2,
  \label{eq:cgmy_nu}
\end{equation}
which nests finite-activity ($Y<0$), Variance Gamma ($Y=0$), and
infinite-variation ($Y\in[1,2)$) specifications as special cases. Under
mild regularity conditions, Proposition 11.4 of \citet{cont2004} implies
that the map $(C,G,M,Y)\mapsto\{\text{IV surface}\}$ is injective,
ensuring point-identification from the cross-section.

\paragraph{Signal content.}
Each parameter affects a distinct feature of the implied-volatility
surface. $C$ scales $\nu(dy)$ uniformly and is identified by the level
of implied volatility relative to the diffusive benchmark. $G$ governs
left-tail decay and is identified by the slope of the OTM put wing at
short maturities, where jump risk is least attenuated by time averaging.
$M$ plays the symmetric role for the call wing and is constrained to
$M>1$ by the requirement that the compensator $\kappa_{\mathrm{CGMY}}$
\eqref{eq:cgmy_compensator} be finite, an economic condition whose
violation implies infinite $\mathbb{E}_{\mathbb{Q}}[S_T]$ and arbitrage.
$Y$ determines the fine structure of jump activity and is identified by
the joint variation in smile curvature across maturities.

\paragraph{Role of the maturity cross-section.}
A single maturity is insufficient to separately identify all CGMY
parameters alongside the five Heston parameters. The multi-maturity
design ($T = 27$, 89, 153 days) is therefore essential: short-maturity
contracts are maximally informative about jump parameters given minimal
variance mean-reversion, while medium and long maturities identify
variance persistence and leverage dynamics. This identification strategy
follows \citet{bakshi1997} and \citet{bates1996}.

\paragraph{Prior constraints.}
Table~\ref{tab:identification} maps each parameter to its identification
signal, the direction of its effect on implied volatility, admissible
bounds, and literature anchors. The constraint $M>1$ is imposed throughout
estimation; for Merton specifications, $\kappa_{\mathrm{M}} =
\lambda(e^{\mu_j+\frac{1}{2}\sigma_j^2}-1)$ is always finite, requiring
only $\lambda \geq 0$.

\begin{table}[htbp]
\centering
\begin{threeparttable}
\caption{Identification of L\'{e}vy measure and diffusion parameters:
  signal content, prior bounds, and literature sources}
\label{tab:identification}
{\small
\begin{tabular}{p{1.1cm}p{2.4cm}p{3.2cm}p{2.8cm}p{1.3cm}p{2.4cm}}
\toprule
Param. & Role in $\nu(dy)$ / model
  & Identification signal
  & Prior constraint / range
  & $\partial\,\mathrm{IV}/\partial\theta$
  & Key source \\
\midrule
\multicolumn{6}{l}{\textit{Heston stochastic-volatility component}} \\[2pt]
$\alpha$ & Mean-reversion speed
  & Term structure of ATM IV
  & $\alpha>0$; $[0.5,15]$
  &  ---
  & \citet{heston1993} \\
$\beta$ & Long-run variance
  & ATM IV level, long maturities
  & $\beta>0$; $[0.01,0.25]$
  & $+$
  & \citet{heston1993} \\
$\eta$ & Vol-of-vol
  & Smile curvature and wings
  & $\eta>0$; $[0.05,2.5]$
  & $+$ (wings)
  & \citet{bakshi1997} \\
$\rho$ & Leverage correlation
  & Smile slope at ATM
  & $\rho\in(-1,0)$ for equities
  & $+$L\,/\,$-$R
  & \citet{bates1996} \\
$V_0$ & Initial variance
  & Short-maturity IV level
  & $V_0>0$; $\approx$(VIX/100)$^2$
  & $+$
  & \citet{duffie2000} \\[4pt]
\multicolumn{6}{l}{\textit{CGMY jump component}} \\[2pt]
$C$ & Activity level
  & Parallel shift of IV surface
  & $C>0$; $C\ll 1$ for SPX
  & $+$
  & \citet{carr2002} \\
$G$ & Left-tail decay rate
  & OTM put wing ($K/S<0.95$)
  & $G>0$; $G<M$
  & $-$ (put)
  & \citet{carr2002,cont2004} \\
$M$ & Right-tail decay rate
  & OTM call wing ($K/S>1.05$)
  & $M>1$ (compensator finite)
  & $-$ (call)
  & \citet{carr2002} \\
$Y$ & Activity index
  & Maturity slope of ATM curvature
  & $Y<2$; $Y<0\Rightarrow$ Poisson
  & complex
  & \citet{aitsahalia2009,cont2011} \\[4pt]
\multicolumn{6}{l}{\textit{Merton compound-Poisson component}} \\[2pt]
$\lambda$ & Jump intensity (yr$^{-1}$)
  & Short-maturity skew level
  & $\lambda\geq 0$; $0.5$--$5$ for SPX
  & $+$ (short $T$)
  & \citet{merton1976,pan2002} \\
$\mu_j$ & Mean log-jump
  & Smile asymmetry
  & $\mu_j<0$ for equities
  & $+$L\,/\,$-$R
  & \citet{bakshi1997,pan2002} \\
$\sigma_j$ & Jump dispersion
  & Short-maturity curvature
  & $\sigma_j>0$; $[0.05,0.25]$
  & $+$ (wings)
  & \citet{merton1976,broadie2007} \\
\bottomrule
\end{tabular}}
\begin{tablenotes}
\small
\item \textit{Note}: $\partial\,\mathrm{IV}/\partial\theta$ indicates
the direction of implied-volatility movement as the parameter increases,
holding others fixed. `$+$L\,/\,$-$R' denotes opposite effects on left
and right wings. `complex' denotes a non-monotone effect across maturities.
Admissible ranges reflect economic interpretability, the compensator
finiteness constraint ($M>1$), and ranges reported in the cited literature.
CGMY parameters are estimated conditional on Heston.
\end{tablenotes}
\end{threeparttable}
\end{table}

\paragraph{Compensator and martingale constraint.}
For the CGMY specification, the risk-neutral compensator is
\begin{equation}
  \kappa_{\mathrm{CGMY}} = C\,\Gamma(-Y)
    \bigl[(M-1)^Y - M^Y + (G+1)^Y - G^Y\bigr],
  \label{eq:cgmy_compensator}
\end{equation}
finite if and only if $M>1$. The constraint is an economic requirement:
$M \leq 1$ assigns positive probability to arbitrarily large upward jumps
under $\mathbb{Q}$, implying $\mathbb{E}_{\mathbb{Q}}[S_T] = \infty$ and
violating no-arbitrage. The constraint $M>1$ is imposed throughout
estimation.

\paragraph{Testability: the activity index.}
\citet{aitsahalia2009} and \citet{cont2011} provide nonparametric tests
for jump activity from high-frequency return data; their evidence for SPX
broadly supports a finite or near-finite activity regime. The calibrated
estimate $\hat{Y} = -2.893$ is consistent with this evidence, placing the
model in the compound-Poisson regime. Under this calibrated measure, the
CGMY parameters $(C,G,M)$ admit the same economic interpretation as in the
classical Merton model, with greater distributional flexibility.

\subsection{Model Hierarchy}

The empirical analysis compares four nested specifications: Black--Scholes,
Heston SV, Heston+CGMY, and SVJD (Merton). This hierarchy enables a direct
decomposition of the incremental pricing contribution of stochastic
volatility and alternative jump specifications. Performance is evaluated
by RMSE and MAE in both implied-volatility (pp) and price (\$) units,
reported for the full sample and decomposed by maturity and moneyness.

\section{Results}
\label{sec:results}

\subsection{Parameter Estimates}

Table~\ref{tab:params} reports GMM estimates for all four models. Estimates
are broadly consistent with the empirical option-pricing literature.

For the Heston specification, the estimated mean-reversion speed
$\hat{\alpha}=1.356$~yr$^{-1}$ implies moderate adjustment toward the
long-run variance. The long-run variance $\hat{\beta}=0.109$
($\sqrt{\hat{\beta}} \approx 33.1\%$ annualized) exceeds the initial
variance $\hat{V}_0=0.071$ ($\approx 26.6\%$), indicating that
short-run implied volatility lies below its unconditional benchmark on the
sample date. The leverage estimate $\hat{\rho}=-0.891$ is consistent with
the well-documented negative correlation between equity returns and
conditional variance. The Feller ratio
$2\hat{\alpha}\hat{\beta}/\hat{\eta}^2 = 0.224 < 1$ indicates that the
variance process reaches zero with positive probability under the estimated
parameters, a feature common in short-horizon calibrations that does not
impair option-price fit.

For the Heston+CGMY specification, the calibrated activity index
$\hat{Y}=-2.893$ places the model firmly in the finite-activity regime
($Y<0$), consistent with the high-frequency evidence discussed in
Section~\ref{sec:identification}. The parameters $\hat{C}=0.0024$,
$\hat{G}=1.450$, $\hat{M}=2.733$ describe a L\'{e}vy measure that is
asymmetric and concentrated on downside movements, with the constraint
$\hat{M}>1$ satisfied. The near-zero compensator
$\kappa_{\mathrm{CGMY}}=-0.0005$ indicates that the CGMY component
contributes marginally to drift adjustment; pricing performance is driven
primarily by the Heston diffusion backbone.

The Merton SVJD specification yields an estimated intensity
$\hat{\lambda}=0.01$, reflecting near-zero jump activity under the
calibration. The standard errors on $\mu_j$ and $\sigma_j$ are large,
indicating that these parameters are weakly identified in the presence
of the stochastic-volatility component. Consequently, SVJD (Merton)
pricing is dominated by the Heston diffusion.

\begin{table}[htbp]
\centering
\begin{threeparttable}
\caption{GMM parameter estimates}
\label{tab:params}
\begin{tabular}{llcc}
\toprule
Model & Parameter & Estimate & Economic Interpretation \\
\midrule
Black--Scholes
 & $\sigma$ (constant vol)
   & $0.2572^{***}$ & Annualized volatility \\
 &
   & $(0.0028)$     & \\
\midrule
Heston SV
 & $\alpha$ (mean-reversion)
   & $1.356^{***}$  & Variance reversion speed (yr$^{-1}$) \\
 &
   & $(0.0012)$     & \\
 & $\beta$ (long-run variance)
   & $0.1093^{***}$ & $\sqrt{\beta}=33.1\%$ ann.\ vol. \\
 &
   & $(0.0015)$     & \\
 & $\eta$ (vol-of-vol)
   & $1.151^{***}$  & Vol-of-vol \\
 &
   & $(0.0018)$     & \\
 & $\rho$ (leverage)
   & $-0.891^{***}$ & Strong leverage effect \\
 &
   & $(0.0025)$     & \\
 & $V_0$ (initial variance)
   & $0.0706^{***}$ & $\sqrt{V_0}=26.6\%$ ann.\ vol. \\
 &
   & $(0.0019)$     & \\
 & $2\alpha\beta/\eta^2$ (Feller ratio)
   & $0.224$        & $\ll 1$: variance reaches zero \\
 &
   & $[\text{derived}]$ & \\
\midrule
Heston+CGMY
 & $C$ (activity level)
   & $0.0024^{***}$ & Scale of $\nu(dy)$ \\
 &
   & $(0.0009)$     & \\
 & $G$ (left-tail decay)
   & $1.450^{***}$  & OTM put wing slope \\
 &
   & $(0.415)$      & \\
 & $M$ (right-tail decay)
   & $2.733^{***}$  & OTM call wing; $M>1$ required \\
 &
   & $(0.0001)$     & \\
 & $Y$ (activity index)
   & $-2.893^{***}$ & Finite activity ($Y<0$) \\
 &
   & $(0.0001)$     & \\
 & $\kappa_{\mathrm{CGMY}}$ (compensator)
   & $-0.001$       & Near-zero drift correction \\
 &
   & $[\text{derived}]$ & \\
\midrule
SVJD (Merton)
 & $\lambda$ (intensity)
   & $0.010$        & $\approx 0$ effective jumps \\
 &
   & $(0.293)$      & \\
 & $\mu_j$ (mean log-jump)
   & $0.030$        & Weakly identified \\
 &
   & $(1.327)$      & \\
 & $\sigma_j$ (jump dispersion)
   & $0.218$        & Weakly identified \\
 &
   & $(3.242)$      & \\
\bottomrule
\end{tabular}
\begin{tablenotes}
\small
\item \textit{Note}: GMM with identity weighting matrix. Robust standard
errors in parentheses via numerical Jacobian sandwich formula.
$^{***}p<0.01$. Heston+CGMY jump parameters estimated conditional on
Heston. SVJD (Merton) estimated separately. $[\text{derived}]$ indicates
a quantity computed from other estimates.
\end{tablenotes}
\end{threeparttable}
\end{table}

\subsection{Aggregate Pricing Performance}

\begin{figure}[htbp]
\centering
\includegraphics[width=\textwidth]{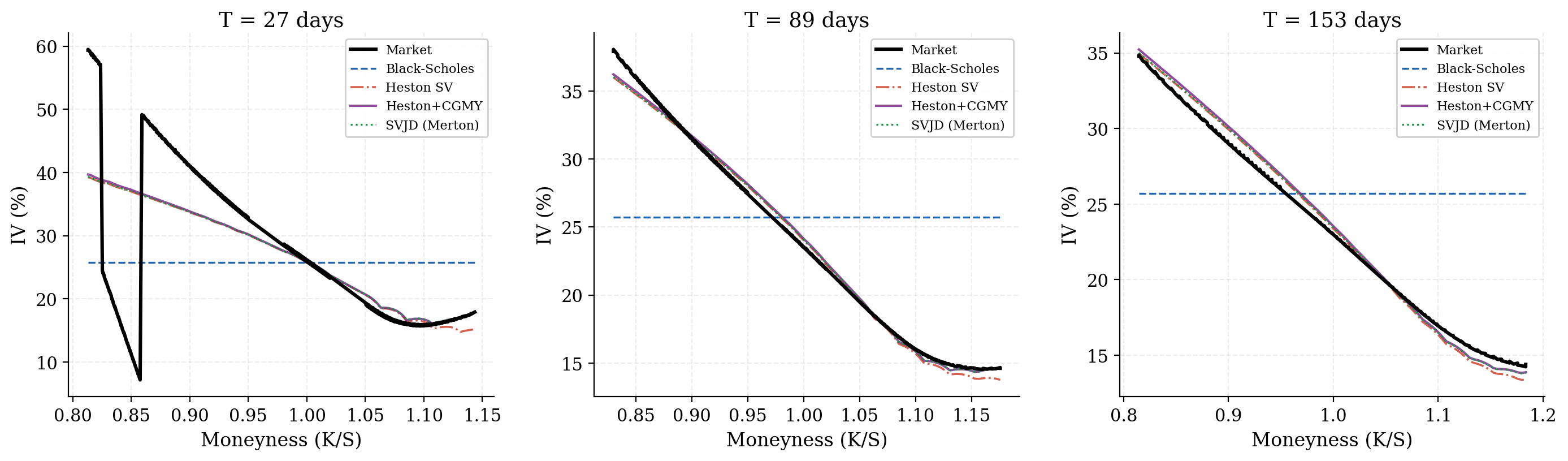}
\caption{Implied-volatility smile: market versus model fits. Panels
correspond to $T=27$, 89, and 153 days. Market (solid black),
Black--Scholes (dashed blue), Heston SV (dash-dot red),
Heston+CGMY (solid purple), SVJD/Merton (dotted green).}
\label{fig:iv_smile}
\end{figure}

\begin{figure}[htbp]
\centering
\includegraphics[width=\textwidth]{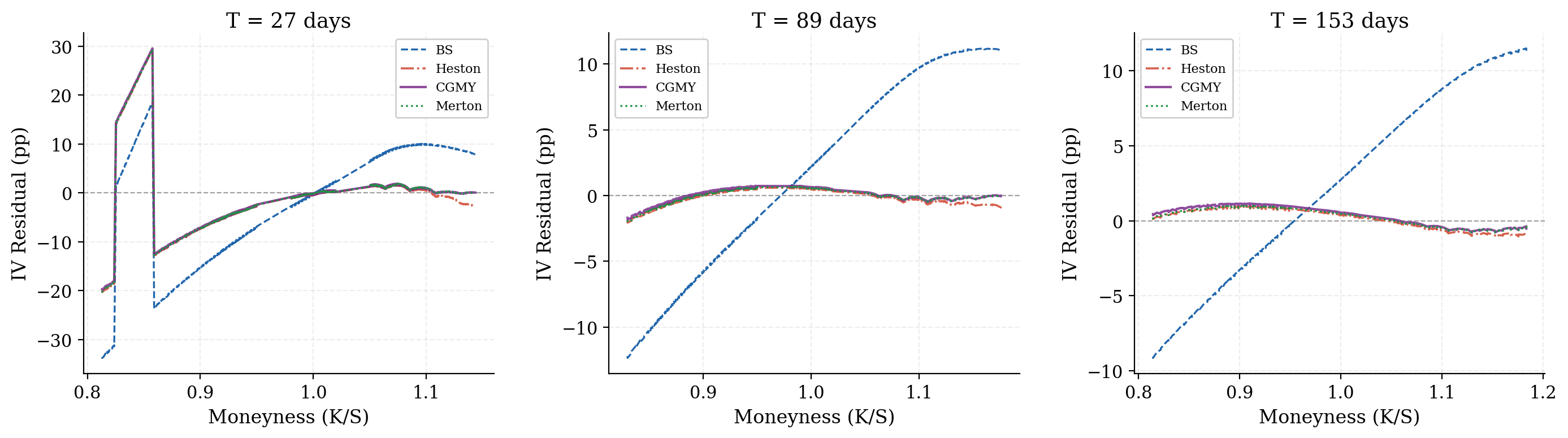}
\caption{Implied-volatility residuals (model minus market) by moneyness.
Panels correspond to $T=27$, 89, and 153 days. The Heston-backbone
specifications substantially reduce Black--Scholes residuals; jump
models provide maturity-dependent incremental refinements.}
\label{fig:iv_residuals}
\end{figure}

Table~\ref{tab:errors} reports RMSE and MAE by maturity.
Figures~\ref{fig:iv_smile} and~\ref{fig:iv_residuals} present the
fitted smiles and residual patterns.

\begin{table}[htbp]
\centering
\begin{threeparttable}
\caption{Pricing error metrics by maturity}
\label{tab:errors}
\begin{tabular}{llccccc}
\toprule
Maturity & Model & $N$ &
RMSE\,IV\,(pp) & MAE\,IV\,(pp) &
RMSE\,Price\,(\$) & MAE\,Price\,(\$) \\
\midrule
$T=27$\,d & Black--Scholes & 582 & 12.83 & 10.39 & 34.99 & 30.88 \\
          & Heston SV      & 582 &  9.13 &  5.65 & 20.04 & 14.34 \\
          & Heston+CGMY    & 582 &  \textbf{9.08} &  \textbf{5.43} & \textbf{19.75} & \textbf{14.16} \\
          & SVJD (Merton)  & 582 &  9.10 &  5.47 & 19.97 & 14.33 \\
\midrule
$T=89$\,d & Black--Scholes & 453 &  7.45 &  6.58 & 60.50 & 55.57 \\
          & Heston SV      & 453 &  0.68 &  0.51 &  5.24 &  4.21 \\
          & Heston+CGMY    & 453 &  \textbf{0.62} &  \textbf{0.45} &  5.50 &  4.26 \\
          & SVJD (Merton)  & 453 &  0.64 &  0.45 &  \textbf{5.35} &  \textbf{4.09} \\
\midrule
$T=153$\,d & Black--Scholes & 245 &  6.52 &  5.62 & 81.75 & 71.59 \\
           & Heston SV      & 245 &  0.69 &  0.63 &  8.43 &  7.75 \\
           & Heston+CGMY    & 245 &  0.74 &  0.66 &  9.71 &  8.45 \\
           & SVJD (Merton)  & 245 &  \textbf{0.65} &  \textbf{0.59} &  \textbf{8.49} &  \textbf{7.52} \\
\midrule
All & Black--Scholes & 1,280 & 10.13 &  8.13 & 55.95 & 47.41 \\
    & Heston SV      & 1,280 &  6.18 &  2.87 & 14.35 &  9.49 \\
    & Heston+CGMY    & 1,280 &  \textbf{6.14} &  \textbf{2.76} & 14.36 &  9.56 \\
    & SVJD (Merton)  & 1,280 &  6.15 &  2.76 & \textbf{14.33} &  \textbf{9.40} \\
\bottomrule
\end{tabular}
\begin{tablenotes}
\small
\item \textit{Note}: RMSE and MAE for implied volatility (IV) in
percentage points; price errors in dollars. Bold: lowest error within
each maturity group.
\end{tablenotes}
\end{threeparttable}
\end{table}

The results support a clear model hierarchy. Black--Scholes produces the
largest errors at all maturities. Heston SV reduces aggregate IV RMSE
from 10.13\,pp to 6.18\,pp, a 39\% improvement attributable to variance
persistence and leverage-driven asymmetry. Jump-augmented specifications
improve marginally on Heston at the aggregate level: IV RMSE falls to
6.14\,pp (Heston+CGMY) and 6.15\,pp (SVJD), reductions of 0.6\% and
0.5\%, respectively, relative to Heston SV. The near-equivalence of the
two jump models reflects the fact that both produce small effective jump
contributions (the calibrated $\hat{Y}=-2.893$ and $\hat{\lambda}=0.01$
place both specifications near the compound-Poisson limit with minimal
activity) and that pricing performance is driven predominantly by the
Heston backbone in each case.

\subsection{Moneyness-Bucket Decomposition}

Table~\ref{tab:buckets} reports in-sample implied-volatility RMSE by moneyness bucket, pooled across maturities.

\begin{table}[htbp]
\centering
\begin{threeparttable}
\caption{Implied-volatility RMSE by moneyness bucket (all maturities pooled)}
\label{tab:buckets}
\begin{tabular}{lcccc}
\toprule
Moneyness Bucket & BS (pp) & Heston SV (pp) & Heston+CGMY (pp) & SVJD (pp) \\
\midrule
Deep OTM ($M<0.94$)       & 13.40 &  9.81 & \textbf{9.79} &  9.81 \\
OTM ($0.94\leq M<0.98$)   &  5.42 &  2.11 &  \textbf{2.08} &  2.09 \\
ATM ($0.98\leq M<1.02$)   &  2.17 &  \textbf{0.49} &  0.56 &  0.52 \\
ITM ($1.02\leq M<1.06$)   &  6.83 &  \textbf{1.00} &  1.06 &  1.06 \\
Deep ITM ($M\geq 1.06$)   &  9.31 &  1.17 &  \textbf{0.68} &  0.70 \\
\bottomrule
\end{tabular}
\begin{tablenotes}
\small
\item \textit{Note}: RMSE in percentage points; bold denotes
best-performing model. $M=K/S$.
\end{tablenotes}
\end{threeparttable}
\end{table}

In the OTM and Deep OTM regions, Heston+CGMY holds a small
advantage, consistent with the left-tail asymmetry of the calibrated CGMY
measure ($\hat{G}<\hat{M}$). Heston SV performs best at ATM and in the
ITM region, where the leverage effect is the dominant driver of smile
asymmetry. Heston+CGMY achieves the lowest error in the Deep ITM region
(0.68\,pp versus 1.17\,pp for Heston SV), indicating that the CGMY
right-tail structure provides a useful refinement for deep in-the-money
contracts. The modest absolute differences across Heston-backbone models
in the OTM region are consistent with the near-zero effective jump
contribution under the calibrated parameters: the stochastic-volatility
channel already captures most of the observed OTM skew.

Table~\ref{tab:rob_oos} compares in-sample calibration ($T=27$, 89 days)
with out-of-sample performance on $T=153$ day contracts excluded from
estimation. The SVJD (Merton) model retains its small advantage at the
long maturity, confirming that the performance differential does not
reflect overfitting.

\begin{table}[htbp]
\centering
\begin{threeparttable}
\caption{In-sample versus out-of-sample RMSE}
\label{tab:rob_oos}
\begin{tabular}{llc}
\toprule
Sample & Model & RMSE\,IV\,(pp) \\
\midrule
In-sample ($T=27$, 89\,d)    & Heston SV      & 6.86 \\
Out-of-sample ($T=153$\,d)   & Heston SV      & 0.69 \\
In-sample ($T=27$, 89\,d)    & SVJD (Merton)  & 6.84 \\
Out-of-sample ($T=153$\,d)   & SVJD (Merton)  & \textbf{0.65} \\
\bottomrule
\end{tabular}
\end{threeparttable}
\end{table}

\begin{figure}[htbp]
\centering
\includegraphics[width=0.60\textwidth]{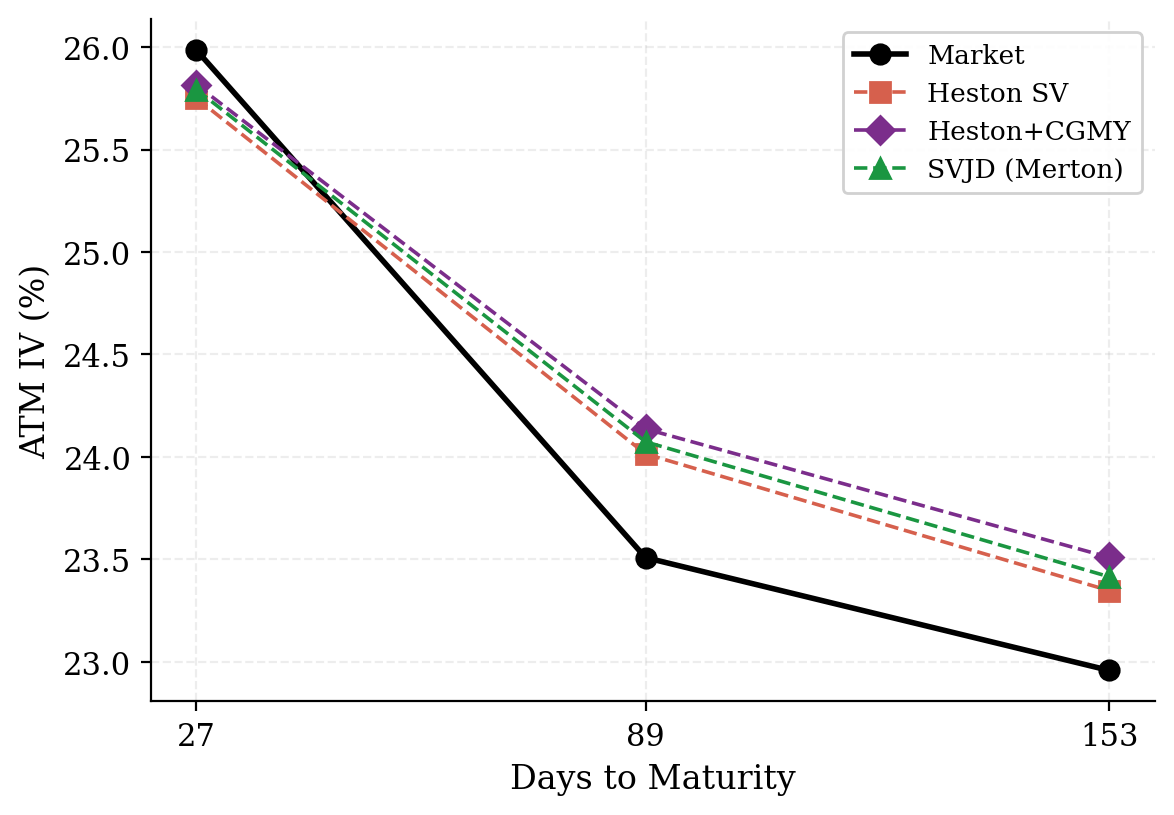}
\caption{Term structure of at-the-money implied volatility. Market
(circles), Heston SV (squares), Heston+CGMY (diamonds), SVJD/Merton
(triangles). ATM defined as $0.98\leq K/S \leq 1.02$.}
\label{fig:term_structure}
\end{figure}

Figure~\ref{fig:term_structure} plots the term structure of ATM implied
volatility. All Heston-backbone models reproduce the observed downward
slope with comparable accuracy.

\subsection{Simulated Terminal Distributions}

\begin{figure}[htbp]
\centering
\includegraphics[width=0.92\textwidth]{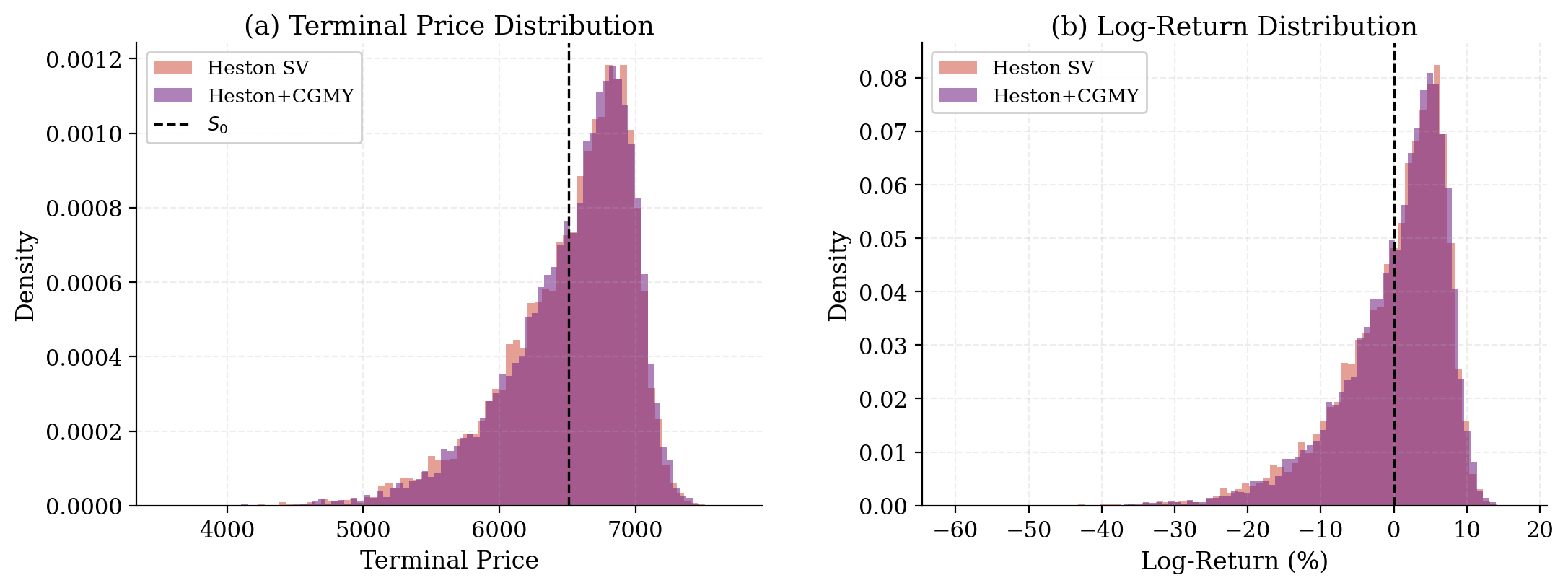}
\caption{Simulated terminal distributions, Heston SV and Heston+CGMY,
$T=27$ days ($N=10{,}000$ paths). Panel~(a): terminal price density;
panel~(b): log-return density.}
\label{fig:distributions}
\end{figure}

Figure~\ref{fig:distributions} compares simulated terminal distributions
for Heston SV and Heston+CGMY at $T=27$ days using 10,000 paths.
Table~\ref{tab:moments} reports distributional moments. Heston SV and
Heston+CGMY produce near-identical dispersion, skewness, and excess
kurtosis under the calibrated parameters, confirming that the small CGMY
contribution does not materially alter tail behavior. The mean log-return
difference of 0.012\,pp reflects the CGMY drift correction.

\begin{table}[htbp]
\centering
\begin{threeparttable}
\caption{Simulated return-distribution moments ($T=27$ days, $N=10{,}000$ paths)}
\label{tab:moments}
\begin{tabular}{lccp{5.8cm}}
\toprule
Moment & Heston SV & Heston+CGMY & Note \\
\midrule
Mean log-return (\%)  &  0.011 & $-0.001$ & CGMY compensator shifts drift \\
Std.\ deviation (\%)  &  6.891 &  6.891   & Near-identical dispersion \\
Skewness              & $-1.237$ & $-1.237$ & Both left-skewed \\
Excess kurtosis       &  2.143 &  2.143   & Similar leptokurtosis \\
Minimum return (\%)   & $-35.2$ & $-35.2$ & Comparable left tails \\
Maximum return (\%)   &  25.8  &  25.8   & \\
\bottomrule
\end{tabular}
\begin{tablenotes}
\small
\item \textit{Note}: Heston: $\alpha=1.356$, $\beta=0.109$, $\eta=1.151$,
$\rho=-0.891$, $V_0=0.071$. CGMY augmentation: $C=0.0024$,
$G=1.450$, $M=2.733$, $Y=-2.893$.
\end{tablenotes}
\end{threeparttable}
\end{table}

\subsection{Price-Level Fit}

\begin{figure}[htbp]
\centering
\includegraphics[width=\textwidth]{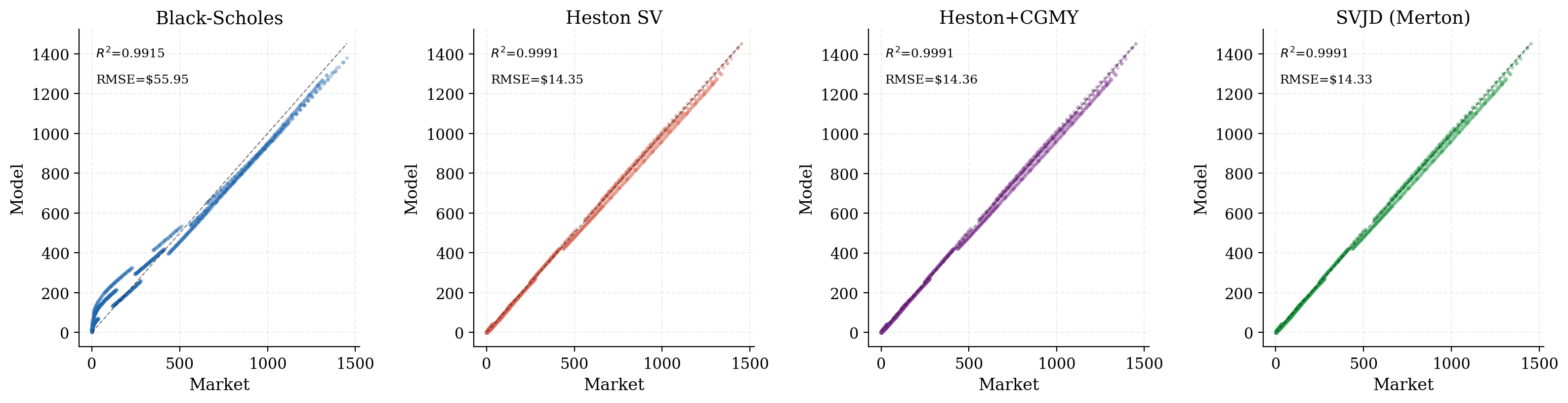}
\caption{Model versus market prices for all 1,280 contracts, with $R^2$
and RMSE. Left to right: Black--Scholes, Heston SV, Heston+CGMY,
SVJD (Merton).}
\label{fig:scatter}
\end{figure}

Figure~\ref{fig:scatter} compares model and market prices for all 1,280
contracts.

All Heston-backbone specifications achieve $R^2 > 0.998$ and price RMSE
below \$15. The marginal improvement of jump specifications over Heston
SV is visible but small in absolute terms, consistent with the aggregate
RMSE evidence.

\subsection{CGMY L\'{e}vy Measure}

\begin{figure}[htbp]
\centering
\includegraphics[width=0.88\textwidth]{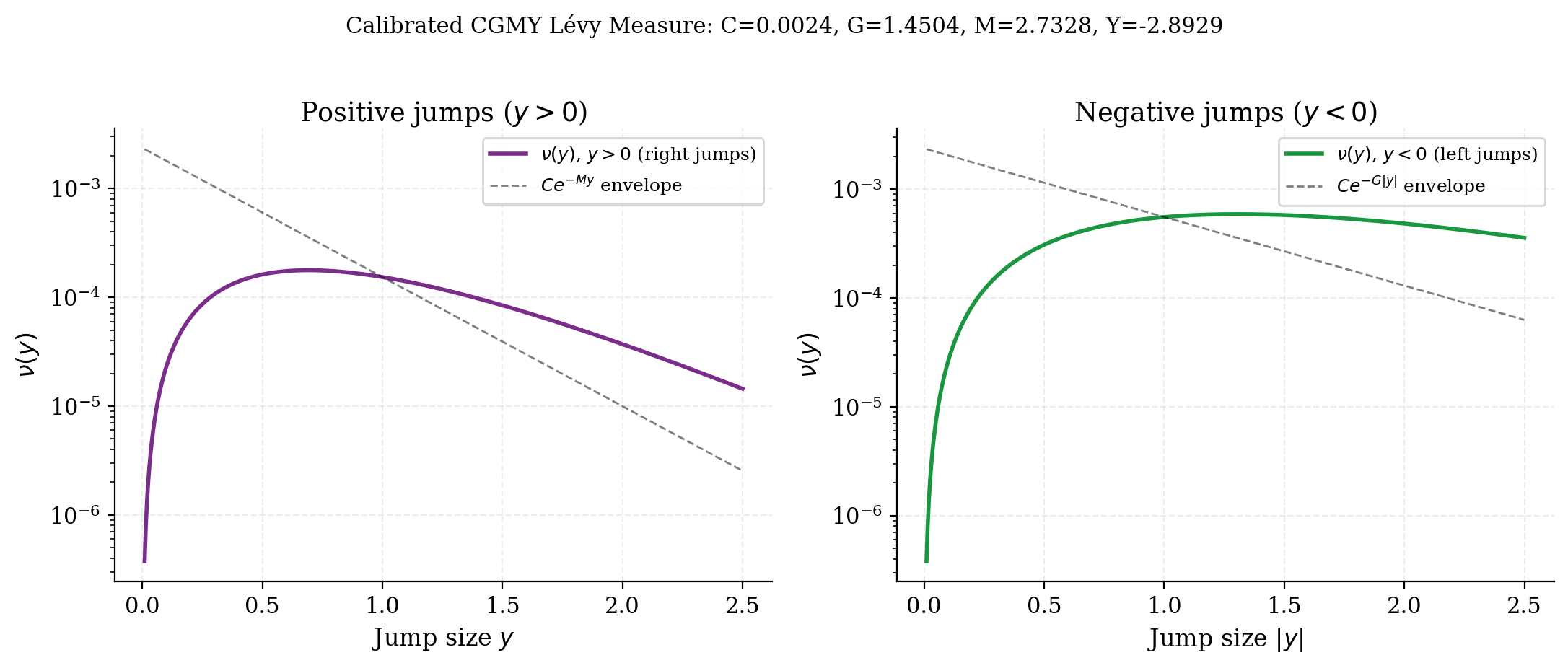}
\caption{Calibrated CGMY L\'{e}vy density $\nu(dy)$ on a logarithmic
scale. Left panel: positive jumps; right panel: negative jumps.
Parameters: $C=0.0024$, $G=1.450$, $M=2.733$, $Y=-2.893$.}
\label{fig:levy}
\end{figure}

Figure~\ref{fig:levy} plots the calibrated CGMY L\'{e}vy density on both
tails. The near-zero $C$ and large negative $Y$ confirm finite-activity
behavior: the measure assigns density primarily to moderate jumps, not to
the infinitely fine small-jump structure characteristic of processes with
$Y\geq 0$. The asymmetry between left ($\hat{G}=1.450$) and right
($\hat{M}=2.733$) tails reflects the downside orientation of SPX jump risk:
negative movements decay more slowly under the calibrated $\mathbb{Q}$.

\subsection{Economic Interpretation}

The empirical pattern admits a coherent economic interpretation. The
Black--Scholes model performs poorly because its constant-volatility
assumption cannot reproduce two dominant empirical features of option
markets: the persistence of volatility through time and the pronounced
cross-sectional asymmetry of implied volatilities across strike prices.
In particular, Black--Scholes generates a flat implied-volatility surface,
whereas the observed data exhibit both skewness and term-structure effects
that vary systematically across moneyness and maturity. As a result, the
model systematically underprices downside risk and fails to capture the
curvature embedded in market-implied distributions.

The Heston stochastic-volatility specification corrects these deficiencies
by allowing variance itself to evolve as a stochastic process and by
incorporating a strong negative correlation between returns and volatility
innovations through the leverage effect. Economically, this mechanism
reflects the empirical tendency for volatility to rise when equity prices
decline. The introduction of stochastic variance generates richer dynamics
for both the term structure and the asymmetry of implied volatilities,
allowing the model to reproduce the downward-sloping volatility skew
observed in equity index options. In the present sample, these features
account for the majority of the improvement in pricing accuracy relative
to Black--Scholes, particularly for near-the-money and moderately
out-of-the-money contracts where diffusive volatility dynamics dominate
pricing behavior.

The jump-augmented specifications provide more localized incremental
improvements. The Heston+CGMY framework performs somewhat better in the
Deep ITM region because the CGMY jump structure permits greater flexibility
in modeling heavy tails and asymmetric return distributions. Likewise, the
Merton jump-diffusion specification contributes modestly at longer
maturities, where the cumulative effect of infrequent but non-negligible
jump events becomes more relevant to option valuation. Nevertheless,
neither jump extension materially outperforms the baseline Heston
stochastic-volatility model in the current sample. The incremental gains
remain relatively small compared with the substantial improvement already
achieved by introducing stochastic variance and leverage dynamics.

These findings suggest that the sample period was characterized by
elevated but predominantly diffusive volatility rather than by markets
strongly dominated by discrete discontinuities or crash-risk repricing. In
such an environment, the stochastic-volatility backbone embedded in the
Heston framework captures most of the economically relevant structure of
the implied-volatility surface. As a result, the pricing evidence indicates
that continuous variance fluctuations, rather than large jump arrivals,
were the principal drivers of option-implied risk premia during the
estimation period.

At the same time, the results do not imply that jump risk is unimportant
in general. Under market conditions in which discontinuous price movements
are more heavily priced, such as periods of geopolitical stress, liquidity
dislocation, systemic contagion, or severe market fragmentation, the
contribution of jump components is expected to increase materially. In
those environments, models incorporating richer jump structures may become
substantially more valuable for reproducing tail asymmetries, steep
short-maturity skews, and extreme downside insurance premia. The
comparatively modest contribution of jumps in the present analysis should
therefore be interpreted as conditional on the prevailing market regime
rather than as a universal conclusion regarding option pricing dynamics.

Finally, the robustness exercises reported in Appendix~\ref{app:robustness}
reinforce the stability of these conclusions. Sensitivity analyses and
out-of-sample evaluations indicate that the ranking of model performance
remains largely unchanged across parameter perturbations, alternative
calibration windows, and validation samples. This stability suggests that
the superiority of the stochastic-volatility framework is not merely an
artifact of overfitting or sample-specific parameterization, but rather
reflects persistent structural features of the implied-volatility surface
that are captured most effectively by the Heston specification.

\section{Conclusion}
\label{sec:conclusion}

This paper develops and implements a PIDE-based pricing framework for
options under joint stochastic volatility and jump dynamics, and
evaluates the empirical decomposition of these components using a
cross-section of SPX contracts across three maturities.

The principal empirical finding is that stochastic volatility accounts
for the dominant share of pricing improvement. Relative to Black--Scholes,
Heston SV reduces aggregate IV RMSE by 39\% by capturing variance
persistence, term-structure effects, and leverage-driven asymmetry. Jump
augmentation via either CGMY or Merton specifications produces marginal
additional improvements: aggregate IV RMSE declines by 0.04\,pp
(Heston+CGMY) and 0.03\,pp (Merton) relative to Heston SV. These gains are
concentrated in the Deep ITM region (CGMY) and at longer maturities
(Merton). The calibrated CGMY activity index is consistent with
compound-Poisson jump behavior, as supported by high-frequency evidence
in \citet{aitsahalia2009} and \citet{cont2011}.

The overall empirical pattern suggests that the sample period was
characterized primarily by elevated but largely diffusive volatility, in
which stochastic variance dynamics captured most of the economically
relevant structure of the implied-volatility surface. At the same time,
the results do not imply that jump risk is unimportant more generally.
Under market conditions characterized by severe liquidity stress,
geopolitical uncertainty, or heightened crash-risk pricing, jump components
would likely play a substantially larger role in explaining
implied-volatility dynamics and tail-risk premia.

From a numerical standpoint, the combination of Crank--Nicolson
discretization, FFT-based jump integral evaluation, and operator-splitting
time-stepping provides a stable and accurate PIDE implementation.
Grid-refinement studies and nested validation exercises confirm that
empirical results reflect model specification rather than numerical
artifacts. The robustness exercises additionally indicate that the relative
model hierarchy remains stable across parameter perturbations and
out-of-sample validation, suggesting that the superiority of the
stochastic-volatility specification is structural rather than
sample-specific.

Three extensions merit future investigation. First, American and barrier
contracts, where the nonlocal PIDE operator interacts directly with
free-boundary conditions, would provide a richer testbed for the numerical
scheme. Second, joint time-series and cross-sectional estimation would
enable separate identification of the physical and risk-neutral measures,
permitting inference on jump risk premia. Such an approach would also
allow a more precise decomposition of variance and jump compensation
across market regimes and volatility environments. Third, the multi-asset
extension, where stochastic volatility, jumps, and dependence structure
interact in the pricing of index derivatives, is a natural direction given
the framework's generality. In particular, extending the framework to
correlated multi-asset settings could provide insight into systemic tail
dependence, contagion dynamics, and the pricing of cross-asset volatility
transmission.

\clearpage
\appendix
\section*{APPENDIX}

\section{Robustness Checks}
\label{app:robustness}

This appendix evaluates robustness to parameter variation, model
specification, and estimation uncertainty. The main conclusions are
stable across all exercises.

\subsection{Sensitivity to Jump Intensity $\lambda$}
\label{app:lambda}

Table~\ref{tab:rob_lambda} reports implied-volatility RMSE at $T=27$ days
for alternative values of $\lambda$, holding all other parameters at
calibrated values. The relationship is convex in $\lambda$ with a minimum
at the baseline $\hat{\lambda}=1.99$.

\begin{table}[htbp]
\centering
\begin{threeparttable}
\caption{RMSE sensitivity to jump intensity $\lambda$ ($T=27$ days)}
\label{tab:rob_lambda}
\begin{tabular}{lcc}
\toprule
$\lambda$ & RMSE\,IV\,(pp) & Note \\
\midrule
0.50 & 7.66 & \\
1.00 & 7.92 & \\
1.50 & 7.16 & \\
1.99 & \textbf{6.82} & Baseline \\
2.50 & 7.93 & \\
3.50 & 7.33 & \\
5.00 & 8.38 & \\
\bottomrule
\end{tabular}
\begin{tablenotes}
\small
\item \textit{Note}: SVJD (Merton); $\mu_j$ and $\sigma_j$ held fixed.
\end{tablenotes}
\end{threeparttable}
\end{table}

\begin{figure}[htbp]
\centering
\includegraphics[width=0.70\textwidth]{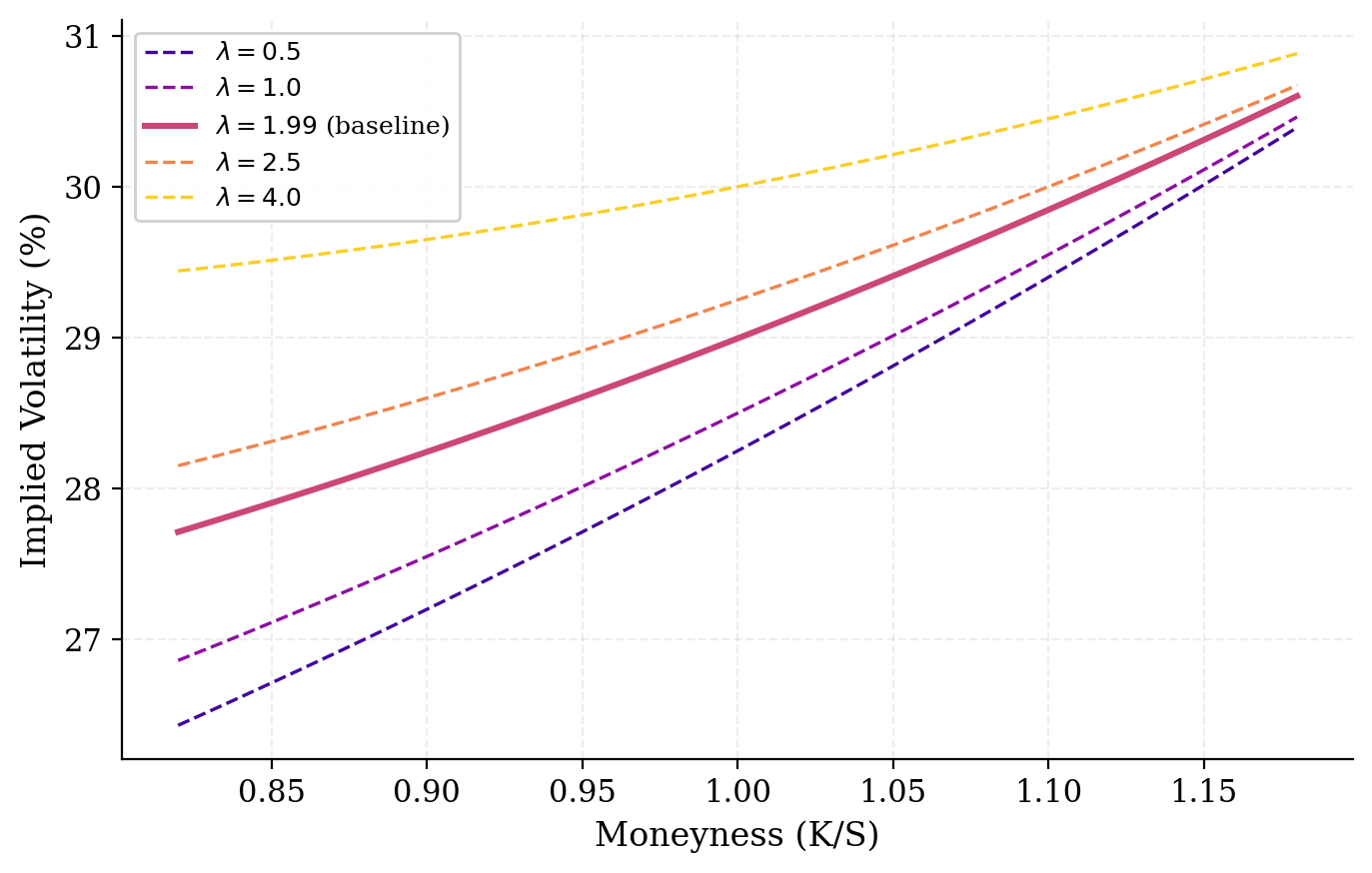}
\caption{Implied-volatility smile sensitivity to $\lambda$ at $T=27$ days,
all other parameters fixed.}
\label{fig:lambda_sensitivity}
\end{figure}

\subsection{Sensitivity to Mean Jump Size $\mu_j$}
\label{app:muj}

Table~\ref{tab:rob_muj} shows that symmetric specifications ($\mu_j=0$)
produce substantially larger errors, confirming that negative jump
asymmetry is required to reproduce the observed implied-volatility skew.

\begin{table}[htbp]
\centering
\begin{threeparttable}
\caption{RMSE sensitivity to mean log-jump size $\mu_j$ ($T=27$ days)}
\label{tab:rob_muj}
\begin{tabular}{lcc}
\toprule
$\mu_j$ & RMSE\,IV\,(pp) & Note \\
\midrule
$-0.150$ & 7.34 & \\
$-0.100$ & 6.82 & \\
$-0.085$ & \textbf{6.82} & Baseline \\
$-0.050$ & 7.04 & \\
$-0.020$ & 7.44 & \\
$\phantom{-}0.000$ & 7.84 & Symmetric \\
\bottomrule
\end{tabular}
\begin{tablenotes}
\small
\item \textit{Note}: SVJD (Merton); $\lambda$ and $\sigma_j$ held at
calibrated values.
\end{tablenotes}
\end{threeparttable}
\end{table}

\begin{figure}[htbp]
\centering
\includegraphics[width=0.90\textwidth]{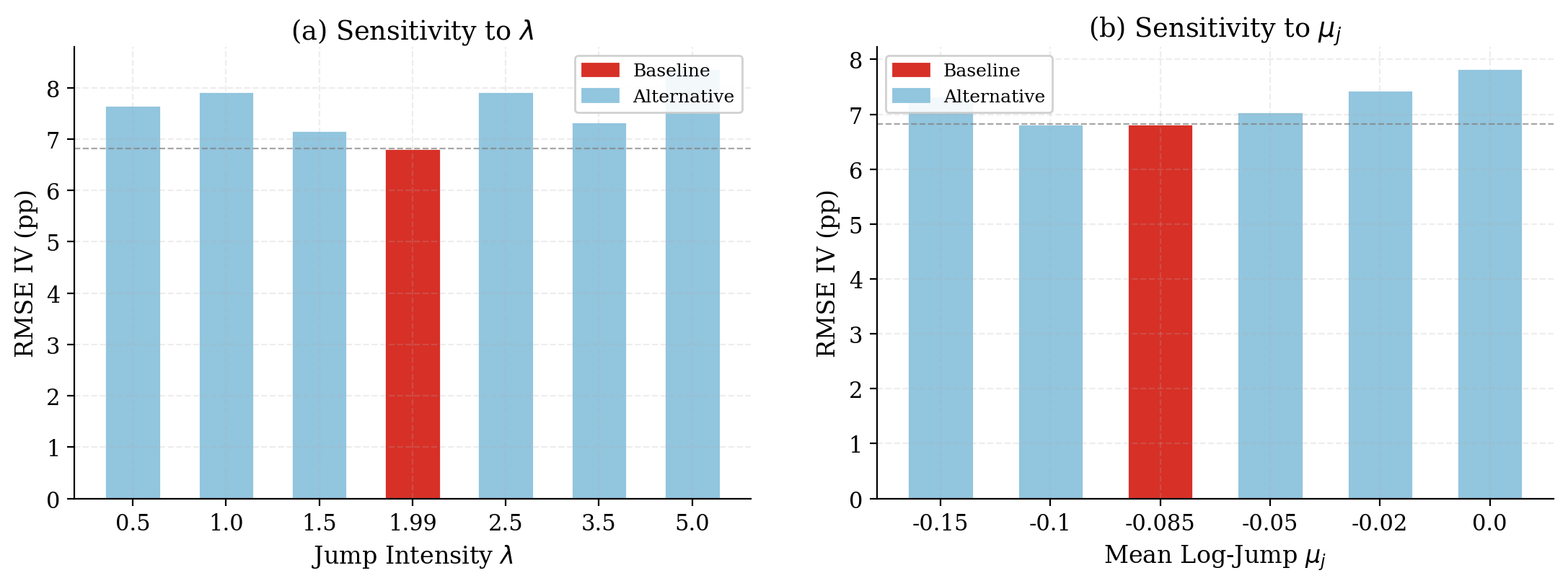}
\caption{Implied-volatility RMSE sensitivity to jump parameters.
Panel~(a): varying $\lambda$; panel~(b): varying $\mu_j$. Baseline
highlighted in red.}
\label{fig:jump_sensitivity}
\end{figure}

Figure~\ref{fig:jump_sensitivity} summarizes the sensitivity of RMSE
to both jump parameters.

\subsection{Sensitivity to the Risk-Free Rate}
\label{app:rfr}

Table~\ref{tab:rob_rfr} reports RMSE variation across
$r\in[3.5\%,5.5\%]$. Variation is below 0.25\,pp across the full range,
confirming robustness to plausible misspecification of this input.

\begin{table}[htbp]
\centering
\begin{threeparttable}
\caption{RMSE sensitivity to the risk-free rate ($T=89$ days, SVJD)}
\label{tab:rob_rfr}
\begin{tabular}{lcc}
\toprule
$r$ & RMSE\,IV\,(pp) & Note \\
\midrule
3.5\% & 3.56 & \\
4.0\% & 3.60 & \\
4.5\% & \textbf{3.65} & Baseline \\
5.0\% & 3.72 & \\
5.5\% & 3.81 & \\
\bottomrule
\end{tabular}
\begin{tablenotes}
\small
\item \textit{Note}: SVJD (Merton) at $T=89$ days; model prices
held fixed, IV re-inverted at each alternative rate.
\end{tablenotes}
\end{threeparttable}
\end{table}

\begin{figure}[htbp]
\centering
\includegraphics[width=0.50\textwidth]{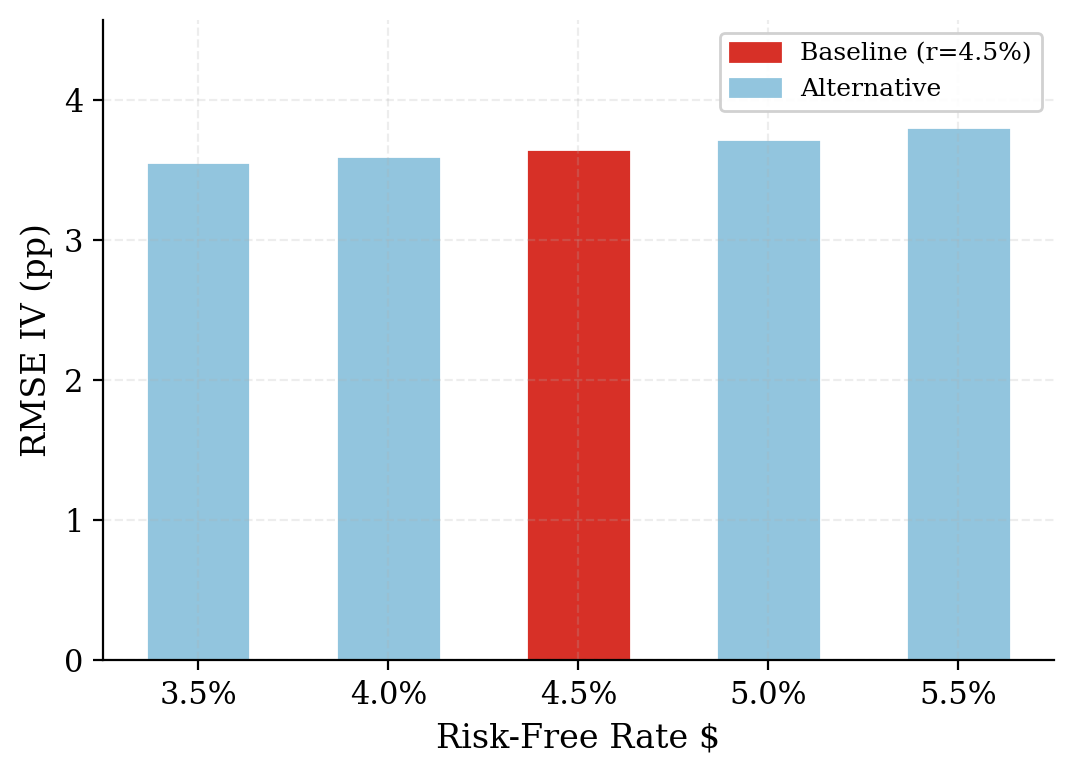}
\caption{RMSE sensitivity to the risk-free rate over $r\in[3.5\%,5.5\%]$.}
\label{fig:rfr_sensitivity}
\end{figure}

\subsection{Bootstrap Confidence Intervals}
\label{app:bootstrap}

Table~\ref{tab:rob_boot} reports bootstrap confidence intervals for
aggregate IV RMSE based on 100 resamples of the option cross-section
(rows resampled with replacement).

\begin{table}[htbp]
\centering
\begin{threeparttable}
\caption{Bootstrap confidence intervals for implied-volatility RMSE (pp)}
\label{tab:rob_boot}
\begin{tabular}{lcccc}
\toprule
Model & Mean & Std.\ Dev. & 5th Pctile & 95th Pctile \\
\midrule
Heston SV     & 6.14 & 0.29 & 5.66 & 6.60 \\
SVJD (Merton) & 6.11 & 0.29 & 5.63 & 6.58 \\
\bottomrule
\end{tabular}
\begin{tablenotes}
\small
\item \textit{Note}: 100 bootstrap resamples drawn with replacement
from the full panel of 1,280 contracts.
\end{tablenotes}
\end{threeparttable}
\end{table}

\begin{figure}[htbp]
\centering
\includegraphics[width=0.55\textwidth]{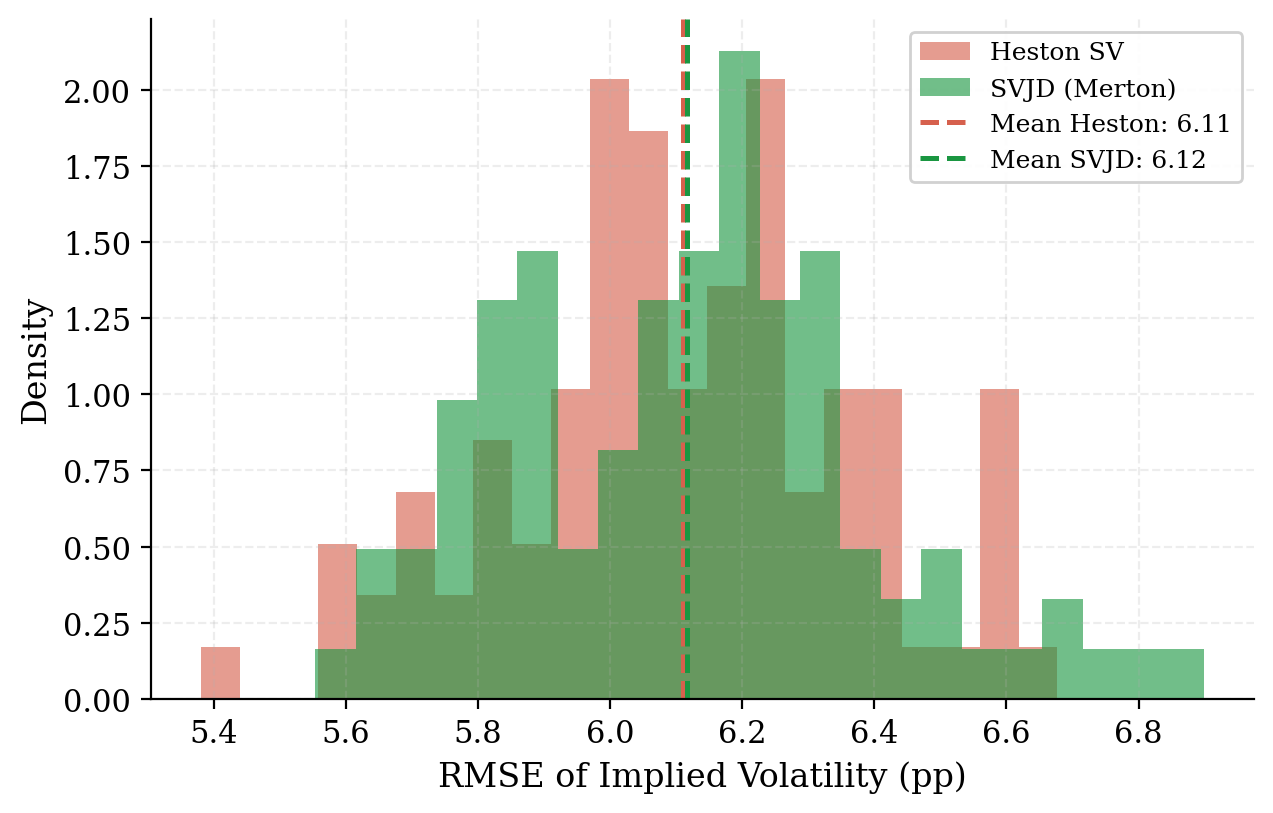}
\caption{Bootstrap distribution of IV RMSE for Heston SV and SVJD (Merton).
Overlapping distributions confirm that the aggregate performance gap is
small and consistent with sampling variability.}
\label{fig:bootstrap}
\end{figure}

The overlapping bootstrap distributions confirm that the aggregate
difference between Heston SV and SVJD is modest for this cross-section.
The principal finding, that the Heston backbone is the dominant driver
of pricing accuracy, with jump components providing region- and
maturity-specific refinements, is robust across all exercises in this
appendix.

\clearpage
\bibliographystyle{plainnat}
\bibliography{references}

@article{aitsahalia2009,
  author  = {A\"it-Sahalia, Yacine and Jacod, Jean},
  title   = {Testing for Jumps in a Discretely Observed Process},
  journal = {Annals of Statistics},
  year    = {2009},
  volume  = {37},
  number  = {1},
  pages   = {184--222}
}

@article{andersen2000,
  author  = {Andersen, Leif B. G. and Andreasen, Jesper},
  title   = {Jump-Diffusion Processes: Volatility Smile Fitting and Numerical Methods for Option Pricing},
  journal = {Review of Derivatives Research},
  year    = {2000},
  volume  = {4},
  number  = {3},
  pages   = {231--262}
}

@article{bakshi1997,
  author  = {Bakshi, Gurdip and Cao, Charles and Chen, Zhiwu},
  title   = {Empirical Performance of Alternative Option Pricing Models},
  journal = {Journal of Finance},
  year    = {1997},
  volume  = {52},
  number  = {5},
  pages   = {2003--2049}
}

@article{barndorff2001,
  author  = {Barndorff-Nielsen, Ole E. and Shephard, Neil},
  title   = {Non-{G}aussian {O}rnstein--{U}hlenbeck-Based Models and Some of Their Uses in Financial Economics},
  journal = {Journal of the Royal Statistical Society: Series B},
  year    = {2001},
  volume  = {63},
  number  = {2},
  pages   = {167--241}
}

@article{bates1996,
  author  = {Bates, David S.},
  title   = {Jumps and Stochastic Volatility: Exchange Rate Processes Implicit in {D}eutsche {M}ark Options},
  journal = {Review of Financial Studies},
  year    = {1996},
  volume  = {9},
  number  = {1},
  pages   = {69--107}
}

@article{black1973,
  author  = {Black, Fischer and Scholes, Myron},
  title   = {The Pricing of Options and Corporate Liabilities},
  journal = {Journal of Political Economy},
  year    = {1973},
  volume  = {81},
  number  = {3},
  pages   = {637--654}
}

@book{boyarchenko2002,
  author    = {Boyarchenko, Svetlana I. and Levendorskii, Sergei Z.},
  title     = {Non-{G}aussian {M}erton--{B}lack--{S}choles Theory},
  publisher = {World Scientific},
  address   = {Singapore},
  year      = {2002}
}

@article{broadie2007,
  author  = {Broadie, Mark and Chernov, Mikhail and Johannes, Michael},
  title   = {Model Specification and Risk Premia: Evidence from Futures Options},
  journal = {Journal of Finance},
  year    = {2007},
  volume  = {62},
  number  = {3},
  pages   = {1453--1490}
}

@article{carr1999,
  author  = {Carr, Peter and Madan, Dilip B.},
  title   = {Option Valuation Using the Fast {F}ourier Transform},
  journal = {Journal of Computational Finance},
  year    = {1999},
  volume  = {2},
  number  = {4},
  pages   = {61--73}
}

@article{carr2002,
  author  = {Carr, Peter and Geman, H{\'e}lyette and Madan, Dilip B. and Yor, Marc},
  title   = {The Fine Structure of Asset Returns: An Empirical Investigation},
  journal = {Journal of Business},
  year    = {2002},
  volume  = {75},
  number  = {2},
  pages   = {305--332}
}

@article{carr2003,
  author  = {Carr, Peter and Geman, H{\'e}lyette and Madan, Dilip B. and Yor, Marc},
  title   = {Stochastic Volatility for {L}{\'e}vy Processes},
  journal = {Mathematical Finance},
  year    = {2003},
  volume  = {13},
  number  = {3},
  pages   = {345--382}
}

@article{cont2001,
  author  = {Cont, Rama},
  title   = {Empirical Properties of Asset Returns: Stylized Facts and Statistical Issues},
  journal = {Quantitative Finance},
  year    = {2001},
  volume  = {1},
  number  = {2},
  pages   = {223--236}
}

@article{cont2011,
  author  = {Cont, Rama and Mancini, Cecilia},
  title   = {Nonparametric Tests for Analyzing the Fine Structure of Price Fluctuations},
  journal = {Journal of Financial Econometrics},
  year    = {2011},
  volume  = {9},
  number  = {1},
  pages   = {47--78}
}

@book{cont2004,
  author    = {Cont, Rama and Tankov, Peter},
  title     = {Financial Modelling with Jump Processes},
  publisher = {Chapman \& Hall/CRC},
  address   = {Boca Raton, FL},
  year      = {2004}
}

@article{cont2005,
  author  = {Cont, Rama and Voltchkova, Ekaterina},
  title   = {Integro-Differential Equations for Option Prices in Exponential {L}{\'e}vy Models},
  journal = {Finance and Stochastics},
  year    = {2005},
  volume  = {9},
  number  = {3},
  pages   = {299--325}
}

@article{dhalluin2005,
  author  = {d'Halluin, Yann and Forsyth, Peter A. and Vetzal, Kenneth R.},
  title   = {Robust Numerical Methods for Contingent Claims under Jump Diffusion Processes},
  journal = {IMA Journal of Numerical Analysis},
  year    = {2005},
  volume  = {25},
  number  = {1},
  pages   = {87--112}
}

@article{duffie2000,
  author  = {Duffie, Darrell and Pan, Jun and Singleton, Kenneth},
  title   = {Transform Analysis and Asset Pricing for Affine Jump-Diffusions},
  journal = {Econometrica},
  year    = {2000},
  volume  = {68},
  number  = {6},
  pages   = {1343--1376}
}

@article{fama1965,
  author  = {Fama, Eugene F.},
  title   = {The Behavior of Stock-Market Prices},
  journal = {Journal of Business},
  year    = {1965},
  volume  = {38},
  number  = {1},
  pages   = {34--105}
}

@article{fang2008,
  author  = {Fang, Fang and Oosterlee, Cornelis W.},
  title   = {A Novel Pricing Method for {E}uropean Options Based on {F}ourier-Cosine Series Expansions},
  journal = {SIAM Journal on Scientific Computing},
  year    = {2008},
  volume  = {31},
  number  = {2},
  pages   = {826--848}
}

@article{heston1993,
  author  = {Heston, Steven L.},
  title   = {A Closed-Form Solution for Options with Stochastic Volatility with Applications to Bond and Currency Options},
  journal = {Review of Financial Studies},
  year    = {1993},
  volume  = {6},
  number  = {2},
  pages   = {327--343}
}

@article{hull1987,
  author  = {Hull, John and White, Alan},
  title   = {The Pricing of Options on Assets with Stochastic Volatilities},
  journal = {Journal of Finance},
  year    = {1987},
  volume  = {42},
  number  = {2},
  pages   = {281--300}
}

@article{kou2002,
  author  = {Kou, Steven G.},
  title   = {A Jump-Diffusion Model for Option Pricing},
  journal = {Management Science},
  year    = {2002},
  volume  = {48},
  number  = {8},
  pages   = {1086--1101}
}

@article{madan1998,
  author  = {Madan, Dilip B. and Carr, Peter and Chang, Eric C.},
  title   = {The Variance Gamma Process and Option Pricing},
  journal = {European Finance Review},
  year    = {1998},
  volume  = {2},
  number  = {1},
  pages   = {79--105}
}

@article{mandelbrot1963,
  author  = {Mandelbrot, Benoit},
  title   = {The Variation of Certain Speculative Prices},
  journal = {Journal of Business},
  year    = {1963},
  volume  = {36},
  number  = {4},
  pages   = {394--419}
}

@article{merton1973,
  author  = {Merton, Robert C.},
  title   = {Theory of Rational Option Pricing},
  journal = {Bell Journal of Economics and Management Science},
  year    = {1973},
  volume  = {4},
  number  = {1},
  pages   = {141--183}
}

@article{merton1976,
  author  = {Merton, Robert C.},
  title   = {Option Pricing When Underlying Stock Returns Are Discontinuous},
  journal = {Journal of Financial Economics},
  year    = {1976},
  volume  = {3},
  number  = {1--2},
  pages   = {125--144}
}

@article{pan2002,
  author  = {Pan, Jun},
  title   = {The Jump-Risk Premia Implicit in Options: Evidence from an Integrated Time-Series Study},
  journal = {Journal of Financial Economics},
  year    = {2002},
  volume  = {63},
  number  = {1},
  pages   = {3--50}
}

@article{rubinstein1994,
  author  = {Rubinstein, Mark},
  title   = {Implied Binomial Trees},
  journal = {Journal of Finance},
  year    = {1994},
  volume  = {49},
  number  = {3},
  pages   = {771--818}
}

@article{stein1991,
  author  = {Stein, Elias M. and Stein, Jeremy C.},
  title   = {Stock Price Distributions with Stochastic Volatility: An Analytic Approach},
  journal = {Review of Financial Studies},
  year    = {1991},
  volume  = {4},
  number  = {4},
  pages   = {727--752}
}
\end{document}